 \documentclass[twocolumn]{aastex631}
\usepackage{amsmath}

\font\manual=manfnt at 7pt \def\dbend{\hbox{\raise0.9ex\hbox{\manual\char127\hspace{0.6em}}}}

\providecommand{\e}[1]{\ensuremath{\times 10^{#1}}}

\newcounter{INTERNALionstage}

\def\gtsim{\mathrel{\hbox{\rlap{\hbox{\lower4pt\hbox{$\sim$}}}\hbox{$>$}}}}
\def\lesssim{\mathrel{\hbox{\rlap{\hbox{\lower4pt\hbox{$\sim$}}}\hbox{$<$}}}}

\def\A{{\rm\thinspace \AA}}

\def\g{{\rm\thinspace g}}

\def\K{{\rm\thinspace K}}

\def\m{{\rm\thinspace m}}

\def\s{{\rm\thinspace s}}

\def\W{{\rm\thinspace W}}

\def\h0{\mbox{{\rm H}$^0$}}
\DeclareMathAlphabet{\vib}{OML}{cmm}{m}{it}
\newcommand*{\satellite}[1]{\textit{#1}}
\newcommand*{\xmm}{\satellite{XMM-Newton}}
\newcommand*{\chandra}{\satellite{Chandra}}

\renewcommand{\thefootnote}{\fnsymbol{footnote}}

\shorttitle{Gas sloshing and cold fronts in A98}
\graphicspath{{./}{figures/}}

\begin{document}

\title{Gas sloshing and cold fronts in pre-merging galaxy cluster Abell~98}

\author[0000-0002-5222-1337]{Arnab Sarkar}
\affiliation{University of Kentucky, 505 Rose street, Lexington, KY 40506, USA}
\affiliation{Center for Astrophysics $\vert$ Harvard \& Smithsonian, Cambridge, MA 02138, USA}
\affiliation{
Kavli Institute for Astrophysics and Space Research, 
Massachusetts Institute of Technology,
77 Massachusetts Ave, Cambridge, MA 02139, USA, 
Cambridge, MA 02138, USA}
\email{arnabsar@mit.edu}

\author[0000-0002-3984-4337]{Scott Randall}
\affiliation{Center for Astrophysics $\vert$ Harvard \& Smithsonian, Cambridge, MA 02138, USA}

\author{Yuanyuan Su}
\affiliation{University of Kentucky, 505 Rose street, Lexington, KY 40506, USA}

\author[0000-0001-9266-6974]{Gabriella E. Alvarez}
\affiliation{Center for Astrophysics $\vert$ Harvard \& Smithsonian, Cambridge, MA 02138, USA}

\author{Craig L. Sarazin}
\affiliation{University of Virginia, Charlottesville, VA 22904, USA}

\author{Christine Jones}
\affiliation{Center for Astrophysics $\vert$ Harvard \& Smithsonian, Cambridge, MA 02138, USA}

\author{Elizabeth Blanton}
\affiliation{Boston University, Boston, MA 02215, USA}

\author{Paul Nulsen}
\affiliation{Center for Astrophysics $\vert$ Harvard \& Smithsonian, Cambridge, MA 02138, USA}
\affiliation{ICRAR, University of Western Australia, 35 Stirling Hwy, Crawley, WA 6009, Australia}

\author{Priyanka Chakraborty}
\affiliation{Center for Astrophysics $\vert$ Harvard \& Smithsonian, Cambridge, MA 02138, USA}


\author{Esra Bulbul}
\affiliation{Max Planck Institute for Extraterrestrial Physics, Giessenbachstrasse 1, 85748 Garching, Germany}

\author{John Zuhone}
\affiliation{Center for Astrophysics $\vert$ Harvard \& Smithsonian, Cambridge, MA 02138, USA}

\author{Felipe Andrade-Santos}
\affiliation{Center for Astrophysics $\vert$ Harvard \& Smithsonian, Cambridge, MA 02138, USA}

\author{Ryan E. Johnson}
\affiliation{Gettysburg College, Gettysburg, PA 17325, USA}











\begin{abstract}
We present deep {\chandra} observations of 
the pre-merger galaxy cluster 
Abell~98. Abell~98 is a complex
merging system. While the northern
(A98N) and central subclusters (A98S) are
merging along the north-south direction,
A98S is undergoing a separate 
late-stage merger, with two distinct X-ray cores.
We report detection of gas sloshing 
spirals in A98N and in the eastern core 
of A98S. 
We detect two cold front edges
in A98N.
We find two more surface brightness edges 
along the east direction of the eastern 
core and west direction of the 
western core of A98S. 
We measure the temperatures and gas 
densities across those edges, 
and find that the eastern edge appears to be a cold front
while the western edge is a shock front with
a Mach number of $\cal{M}$ $\approx$ 1.5.
We detect a ``tail" of X-ray emission associated with 
the eastern core of A98S. 
Our measurement indicates that
the tail is 
cooler than the surrounding gas at 
a 4.2-$\sigma$ level, 
suggesting the tail is part of a cool 
core remnant that has been ram-pressure stripped.
\end{abstract}



\section{Introduction } \label{sec:intro}

Abell 98 (hereafter A98) is an early-stage 
merger \citep{1981ApJ...243L.133F} 
with three main sub-clusters, 
A98S (central), A98N (northern), and A98SS (southern). 
Previous studies have reported a shock front 
associated with the northern sub-cluster and 
a filament connecting the northern and 
central sub-clusters.  \citep[e.g.,][]{2022ApJ...938...51A,2022ApJ...935L..23S}. 
\citet{2022ApJ...938...51A} also reported a 
filament extending to the north beyond A98N, indicating
A98N and A98S lie along a large-scale cosmic filament.
In this work, we focus on the 
northern and central subclusters. 

Cluster mergers along large scale 
filaments leave imprints on the intra-cluster 
medium (ICM) in the form of shocks, cold fronts, 
and gas sloshing spirals
\citep[e.g.,][]{2001ApJ...562L.153M,2002ApJ...567L..27M,2006ApJ...650..102A,2007PhR...443....1M}.
Numerical simulations show that when a 
subhalo passes the density peak of the 
central cluster 
with a nonzero impact parameter, 
it accelerates the DM and gas peaks towards it.  
At first, they travel together, 
but as the gas peak moves through the surrounding gas, 
the ICM
velocity field around the main cluster core 
decelerates the gas peak due to the ram pressure, 
resulting in a separation between the DM and gas peaks.
While the DM peak continues moving towards the
receding subhalo, the gas peak is held back.
During the core passage, the direction of this motion 
rapidly changes, 
leading to an abrupt change in the ram-pressure 
force exerted on the gas peak, 
which then overshoots the potential minimum. 
Eventually, the gas peak turns around and starts sloshing 
back and forth around the DM peak. 
This sloshing motion can be inferred from 
imaging studies, which reveal the presence of
sharp discontinuities in
the surface brightness and temperature.
In these features,
the brighter (denser) side of the surface
brightness jump is colder than the 
fainter (less dense) side; 
hence this is a ``cold front",
as opposed to the shock front
\citep[e.g.,][]{2000ApJ...541..542M,2001ApJ...551..160V}.

Gas sloshing motion can induce multiple cold fronts
at different radii, 
growing over time and combining into a spiral pattern, 
known as a ``gas sloshing" spiral
\citep[e.g.,][]{2006ApJ...650..102A,2007PhR...443....1M,2011MNRAS.413.2057R}. 
These spiral structures are commonly observed in  
several galaxy clusters, 
such as A1763 \citep{2018ApJ...868..121D},
A2029 \citep{2013ApJ...773..114P},
A2052 \citep{2011ApJ...737...99B},
and the Fornax cluster 
\citep{2017ApJ...851...69S}. 
Measurements of gas properties across the cold 
fronts reveal that the gas pressure 
stays almost the same 
across these cold fronts,
as observed in 
several galaxy clusters, 
e.g., A2029 \citep{2004ApJ...616..178C},
Perseus \citep{2019MNRAS.483.1744I},
A2142 \citep{2013A&A...556A..44R}, and
in galaxy groups \citep{2013ApJ...770...56G}. 
This limits the 
relative gas motion, as ram pressure forces 
would increase the gas pressure inside of the cold front. 
Though there is pressure equilibrium across the cold fronts, 
observations of other clusters show that the gas outside 
the edge is nearly at hydrostatic equilibrium, 
but the gas inside the edge is not 
\citep[e.g.,][]{2001MNRAS.321L..33F,2007PhR...443....1M}.

A98 is an early-stage merger 
at a redshift of $z\approx$ 0.1042 \citep{2014ApJ...791..104P}. 
Previous observations confirm that 
this is a system of Bautz-Morgan type II-III and 
richness of class 3 \citep{1989ApJS...70....1A}. 
The central sub-cluster A98S 
is at a redshift of $z\approx$ 0.1063 and hosts a wide-angle tail (WAT) 
radio source \citep[4C 20.04;][]{1993ApJ...408..428O}. 
The other two sub-clusters, A98N ($z\approx$ 0.1043) 
and A98SS ($z\approx$ 0.1218), are at
projected distances of $\sim$ 1.1 Mpc north 
and 1.4 Mpc south from the central sub-cluster, respectively. 
\citep[e.g.,][]{1999ApJ...511...65J,1994ApJ...423...94B}. 
A98N is a cool core cluster with an asymmetric 
warmer gas arc in the south
\citep{2014ApJ...791..104P}. 
In paper I \citep{2022ApJ...935L..23S}, 
we investigated this scenario in detail. 
We reported a shock front with a Mach number of 
$\cal{M}$ $\approx$ 2.3 at about 420 kpc south
from the A98N core. 

The primary goals of this paper are to map 
the thermodynamic structure 
and reveal any sub-structures 
associated with the central regions of A98N and A98S. 
We investigate the surface brightness edges 
near A98N and A98S and 
measure gas properties across them. 
We organize the paper as follows: in Section \ref{sec:data},
we describe the 
data reduction procedures; in Section \ref{sec:imaging},
we analyse the actual, residual, the
Gaussian Gradient Magnitude (GGM) images, and
report the 
surface brightness profiles and electron density 
jumps;
in Section \ref{sec:spectral}, we show
results from the 
spectral analysis, including the gas properties
profiles; finally, in Section \ref{sec:diss}
we discuss and 
summarize our results.

Throughout this paper, we adopt a cosmology of
H$_0$ = 70 km s$^{-1}$ Mpc$^{-1}$, 
$\Omega_{\Lambda}$ = 0.7, 
and $\Omega_\textrm{m}$ = 0.3.
Unless otherwise stated, all reported error bars 
correspond to a 90\% confidence level.

\begin{deluxetable}{ccccccc}
\tablenum{1}
\tablecaption{$\chandra$ observation log} \label{tab:obs_log}
\tablewidth{0pt}
\tablehead{
\colhead{Obs ID} & \colhead{Obs Date} & \colhead{Exp. time} & \colhead{PI} \\
\colhead{} & \colhead{(J2000)} & \colhead{(ks)} & \colhead{} \\
}
\startdata
11876 & 2009 Sep 17 & 19.2 & S. Murray \\
11877 & 2009 Sep 17 & 17.9 & S. Murray \\
21534 & 2018 Sep 28 & 29.5 & S. Randall \\
21535 & 2019 Feb 19 & 24.7 & S. Randall \\
21856 & 2018 Sep 26 & 30.5 & S. Randall \\
21857 & 2018 Sep 30 & 30.6 & S. Randall \\
21880 & 2018 Oct 09 & 9.9 & S. Randall \\
21893 & 2018 Nov 11 & 17.9 & S. Randall \\
21894 & 2018 Nov 14 & 17.8 & S. Randall \\
21895 & 2018 Nov 14 & 28.6 & S. Randall \\
\enddata
\end{deluxetable}

\begin{figure*}
    \centering
    \includegraphics[scale=0.5]{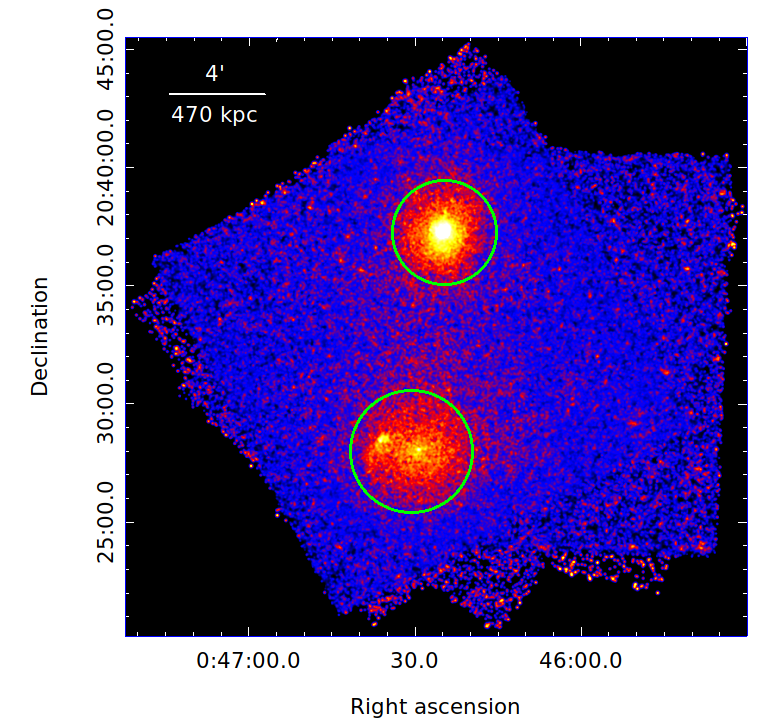}
    \caption{Merged, exposure corrected, and background subtracted
$\chandra$ ACIS-I image of northern (A98N) and 
central (A98S) sub-clusters of A98 in the 0.5--2.0 energy band.
The image has been smoothed with a $\sigma$=2$\arcsec$ Gaussian kernel.
The dual structure of A98S is clearly visible. 
Green circles represent the regions used for
the measurement of global ICM properties of both
sub-clusters.}
    \label{fig:whle_img}
\end{figure*}

\section{Data preparation} \label{sec:data}
Abell 98 was observed with $\chandra$ during two epochs,
once in September 2009 for 37 ks split into two observations 
and later in September 2018 -- February 2019 for 190 ks divided 
into eight observations. 
The combined exposure time is 
$\sim$ 227 ks (detailed observation
logs are listed in Table \ref{tab:obs_log}). 
We performed the $\chandra$ data reduction with CIAO version 
4.12 and CALDB version 4.9.4 distributed by the Chandra X-ray Center (CXC). 
We have followed a standard data analyzing thread
\footnote[4]{\url{http://cxc.harvard.edu/ciao/threads/index.html}}. 

All level 1 event files were reprocessed using the 
{\tt chandra$\_$repro} task by employing the latest gain, charge transfer 
inefficiency correction, and filtering out the bad grades. 
VFAINT mode was used to improve the background screening. 
Light curves were extracted and filtered using the {\tt lc$\_$clean} script to 
identify and remove periods affected by flares. The filtered exposure 
times are listed in Table \ref{tab:obs_log}. 
We used the {\tt repoject$\_$obs} task to reproject 
all observations to a common tangent position and combine them. 
The exposure maps in the 0.5-2.0 keV energy
bands were created using 
the {\tt flux$\_$obs} script by providing a weighting spectrum,
which was generated using the {\tt make$\_$instmap$\_$weights} 
task with an absorbed APEC plasma emission model and  
a plasma temperature of 3 keV. 
To remove the underexposed edges of the
detector,
we set the pixel value to zero for those
pixels with an exposure of 
less than 15\% of the combined exposure time. 

Point sources were identified using {\tt wavdetect} with a range of
wavelet radii between  1--16 pixels. We set the detection 
threshold to $\sim$ 10$^{-6}$, which guaranteed $\lesssim$ 1 spurious source 
detection per CCD. We used blanksky background observations 
to model the non-X-ray 
background, emission from foreground structures 
(e.g., Galactic Halo and Local Hot Bubble) along the observed direction
and unresolved faint background sources. 
The blanksky background files were 
generated using the {\tt blanksky} task and then 
reprojected to match the coordinates of the observations. 
We finally normalized the resulting blanksky background 
to match the 
9.5--12 keV count rates in our observations.

\section{Imaging analysis} \label{sec:imaging}
Figure \ref{fig:whle_img} shows a slightly smoothed image of A98N 
(northern sub-cluster) and A98S (central sub-cluster) 
in the 0.5-2 keV band, obtained after co-adding all ten 
observations and by applying the correction for exposure 
non-uniformity. As reported by earlier studies, 
the cluster is an early-stage merger, with the 
merger axis likely aligned along a local N-S large-scale 
filament connecting 
A98N and A98S \citep[e.g.,][]{2014ApJ...791..104P,2022ApJ...938...51A,2022ApJ...935L..23S}. 

\subsection{A98N} \label{sec:A98N}
\begin{figure*}
\begin{tabular}{cc}
     \hspace{-10pt}
     \includegraphics[scale=0.4]{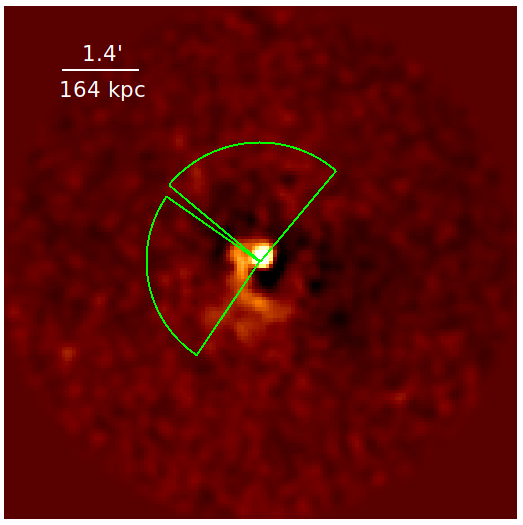} &  
     \includegraphics[scale=0.26]{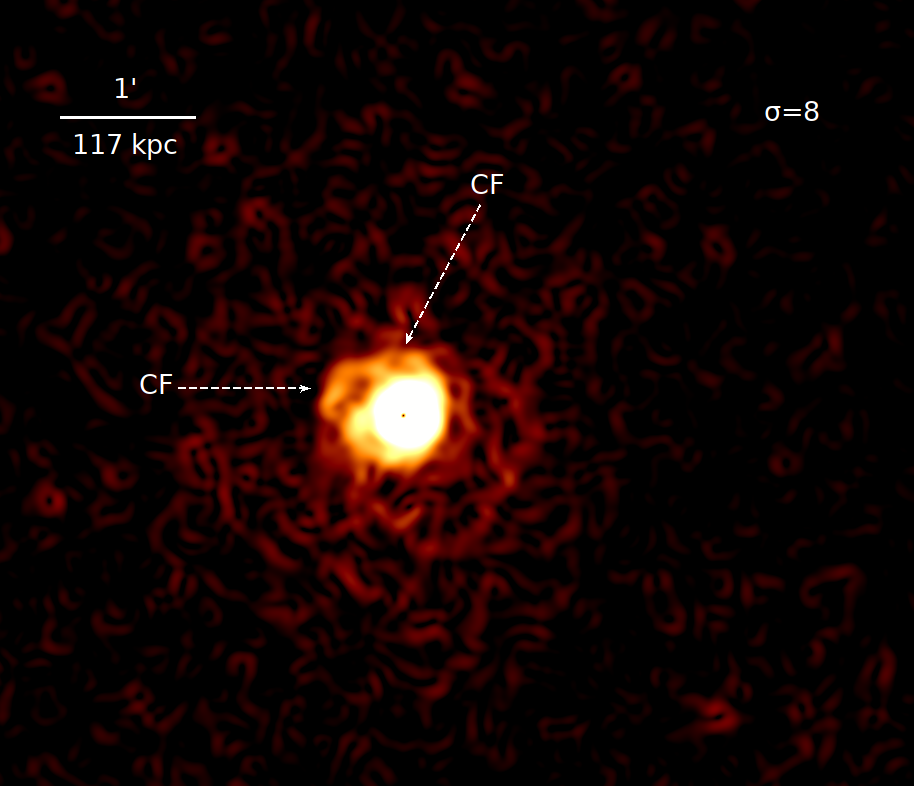}\\
\end{tabular}
\caption{{\it Left}: Residual image of A98N
in the 0.5-2 keV energy band obtained
after subtracting the azimuthal averaged surface brightness profile
from Figure \ref{fig:whle_img}. 
The image has been smoothed with a 1.5$\arcsec$
Gaussian kernel. 
A spiral pattern can clearly be
seen travelling clock-wise if traced inward.
Green sectors are used for spectral analysis.
{\it Right}: Gaussian gradient magnitude filtered image of 
A98N
smoothed by $\sigma$ = 8 pixels and each pixel = 0.492$\arcsec$.
The north and east cold fronts are marked 
by white arrows.
\label{fig:ggm_img}}
\end{figure*}

We used a residual image of A98N to search for any 
faint substructures, e.g., associated with gas sloshing or 
merger shocks. The residual image was created by 
subtracting the azimuthally averaged surface 
brightness at each radius (centered on A98N), as
shown in Figure \ref{fig:ggm_img}. 
The residual image shows an 
apparent surface brightness excess spiraling 
clockwise (if traced inward from large radius), 
a classic signature of gas sloshing, 
and is brightest to the north/north-east 
of the cluster center. The farthest visible arc of the 
spiral extends out to 140 kpc south from the cluster center.
Similar gas sloshing spirals have also been observed in other 
galaxy clusters, e.g., A2029 \citep{2013ApJ...773..114P}, 
Perseus \citep{2011MNRAS.418.2154F}, 
A1763 \citep{2018ApJ...868..121D}, 
A2319 \citep{2021MNRAS.504.2800I},
and
A2142 \citep{2012MNRAS.420.3632R}. 

The sloshing spiral in A98N
indicates A98N experienced 
a separate merger,
un-associated with the on-going 
early-stage merger between A98N and A98S,
probably in the
past few Gyr, based on the timescale 
for the formation of the sloshing
spirals seen in simulations 
\citep{2010ApJ...717..908Z}.
Spiral patterns are understood to have 
formed when a cool-core system experiences an 
infall of a subcluster with a non-zero impact parameter. 
Simulations of the formation of spiral patterns 
show that during the formation process, 
the relative velocity between the gas density and 
dark matter peak stays below 400 km/s \citep{2006ApJ...650..102A}, 
and the average offset between gas density-dark matter peaks 
is below 30 kpc \citep{2010ApJ...710.1776J}. 
Additionally, the winding direction of a spiral 
(when traced inward from large radii) reveals the infalling subcluster's
trajectory, since the direction of angular momentum of 
spiraling-in gas is the same as that of the infalling subcluster 
\citep{2011ApJ...743...16Z,2018ApJ...868..121D}.
The clockwise winding spiral pattern seen in A98N indicates 
that the perturbing subcluster may have passed the A98N core 
from the west to the north-east on the southern side.

\begin{figure*}
\gridline{\hspace{-10pt}
\fig{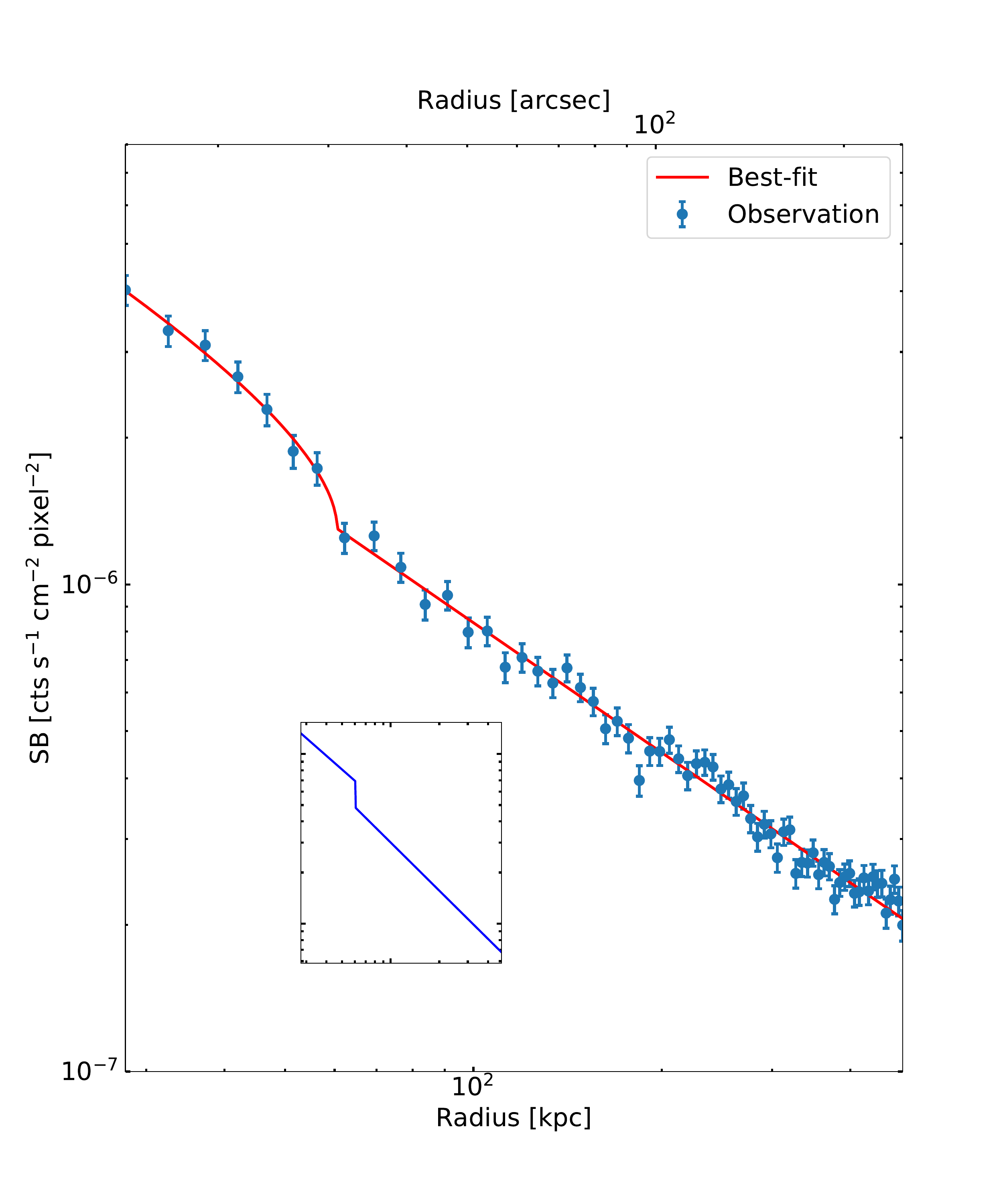}{0.5\textwidth}{a. North}
          \hspace{-20pt}
          \fig{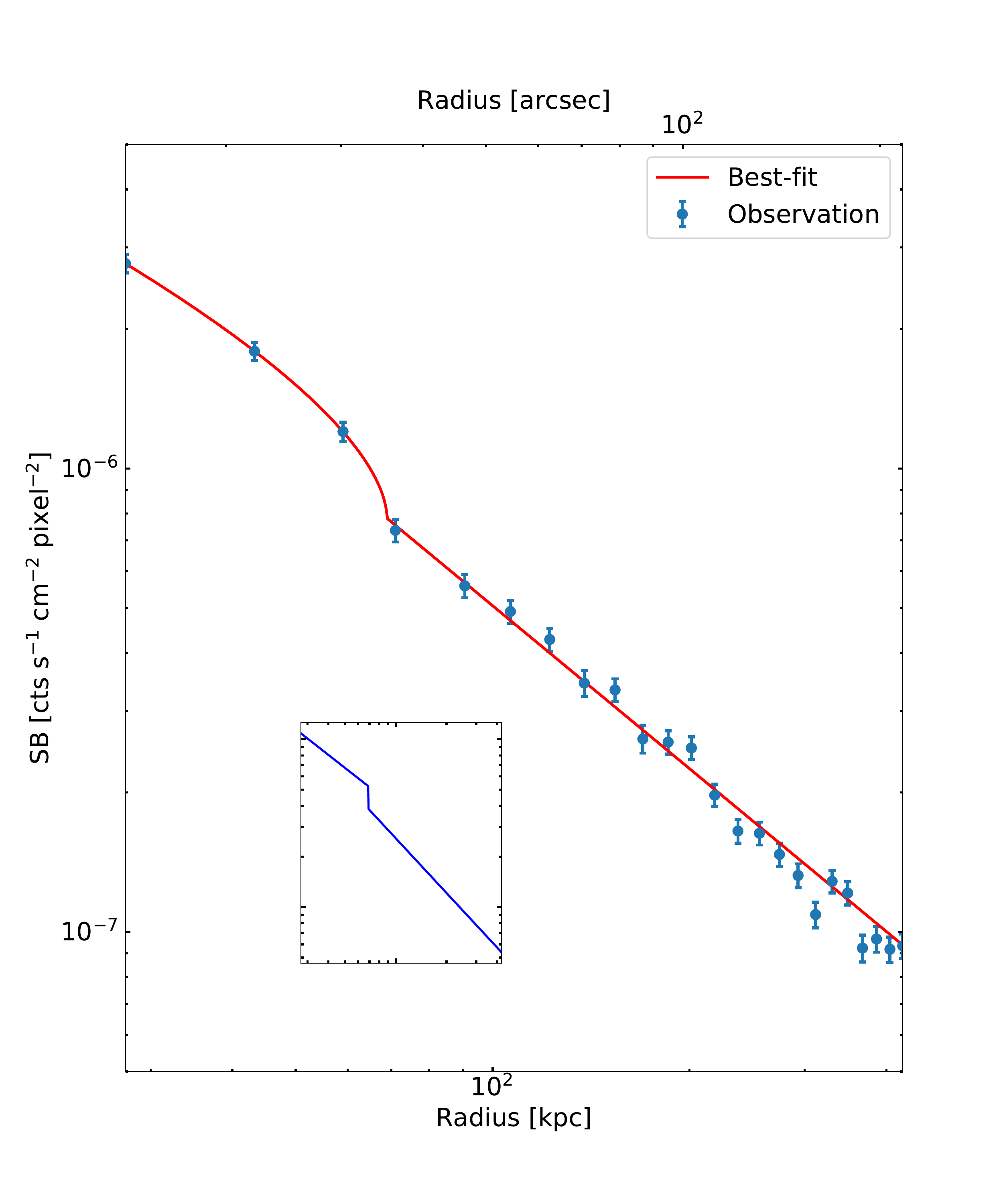}{0.5\textwidth}{b. East}
          }
\caption{{\it Left}: Northern sector surface brightness profile of A98N in the
0.5 - 2 keV energy band fitted with a broken power-law model. The best-fit 
deprojected
density profile is shown in inset figure.
All error bars are at 1-$\sigma$ level.
{\it Right}: similar to the {\it left} figure, but for eastern sector.}
\label{fig:edge_fitting}
\end{figure*}

Next, to reveal any substructures and associated sharp 
features in the A98N core, we obtained a GGM 
filtered image of A98N with 
a filtering length scale of 8 pixels. Each pixel 
is 0.492$\arcsec$ $\times$ 0.492$\arcsec$. 
Figure \ref{fig:ggm_img} (right)
shows the GGM image of the A98N core. 
We observe two regions with relatively 
large surface brightness gradients in the core,
peaking at 
$\sim$ 60 kpc to the north and 70 kpc to the 
east of the A98N core 
(indicated by the white arrows in Figure \ref{fig:ggm_img}), 
sharply declining at larger radii. 
To determine the significance of 
these apparent features, we extracted
surface brightness profiles 
in the east and north directions in 90$^\circ$ sectors. 
The energy band is restricted to 0.5-2 keV to minimize 
the effect of variation of X-ray emissivity with temperature 
while maximizing the signal-to-noise ratio. 
We match the centers of 
the curvature of the surface brightness edges to the center 
of the sectors. Figure \ref{fig:edge_fitting}
shows the extracted surface 
brightness profiles in the east and north directions.

Both surface brightness profiles exhibit discontinuities 
and overall shapes that are consistent 
with what is expected from a projected
spherical density discontinuity
\citep{2000ApJ...541..542M}. 
We fit each profile with a broken power-law model 
\begin{equation}
    n(r) \propto 
    \begin{cases}
    r^{-\alpha_{1}}, & \text{if $r < r_{edge}$}\\
    \frac{1}{jump} r^{-\alpha_{2}}, & \text{if $r \geq r_{edge}$}
    \end{cases}
\end{equation}
where $n(r)$, $r_{edge}$,
and $jump$ 
represent 3D electron density at a radius $r$, 
the radius of the putative edge, and the 
density jump factor, respectively.
$\alpha_1$ and $\alpha_2$
are the slopes inside and outside the edge,
respectively.
We project the estimated emission 
measure profile onto the sky plane and fit the observed surface 
brightness profile by varying the slopes, edge radius, and the 
magnitude of the jump 
(similar to what we did for the 
shock front in paper I; \citealt{2022ApJ...935L..23S}). 
Figure 
\ref{fig:edge_fitting} 
displays the best-fit models (red) and 
the 3D density profiles (inset) for 
both directions. 
For the east direction, we obtain 
best-fit power-law indices of 
$\alpha_1$ = 0.70 $\pm$ 0.01 and 
$\alpha_2$ = 1.0 $\pm$ 0.01
The associated density jump across the edge is 
$\rho_2$/$\rho_1$ = 1.38 $\pm$ 0.03, 
where suffix 2 and 1 represent the regions 
inside and outside of the edge, respectively. 
Similarly, for the north direction, we find a 
best-fit power-law index of 
$\alpha_1$ = 0.80 $\pm$ 0.03 and 
$\alpha_2$ = 1.10 $\pm$ 0.03
with a 
corresponding density jump of 
1.47 $\pm$ 0.01. 
We obtain the best-fit edge radius of 
70 $\pm$ 6 kpc for the 
east edge and 
60 $\pm$ 4 kpc for the north edge,
as measured with respect to the 
centroid of the central 
X-ray surface brightness peak. 
Both edge 
radii are consistent with the position of the 
steep gradients seen in the GGM image (Figure \ref{fig:ggm_img}).
We estimate the uncertainty of each parameter
by allowing 
all the other model parameters to vary freely.

\subsection{A98S}
\begin{figure*}
\begin{tabular}{cc}
     \includegraphics[scale=0.25]{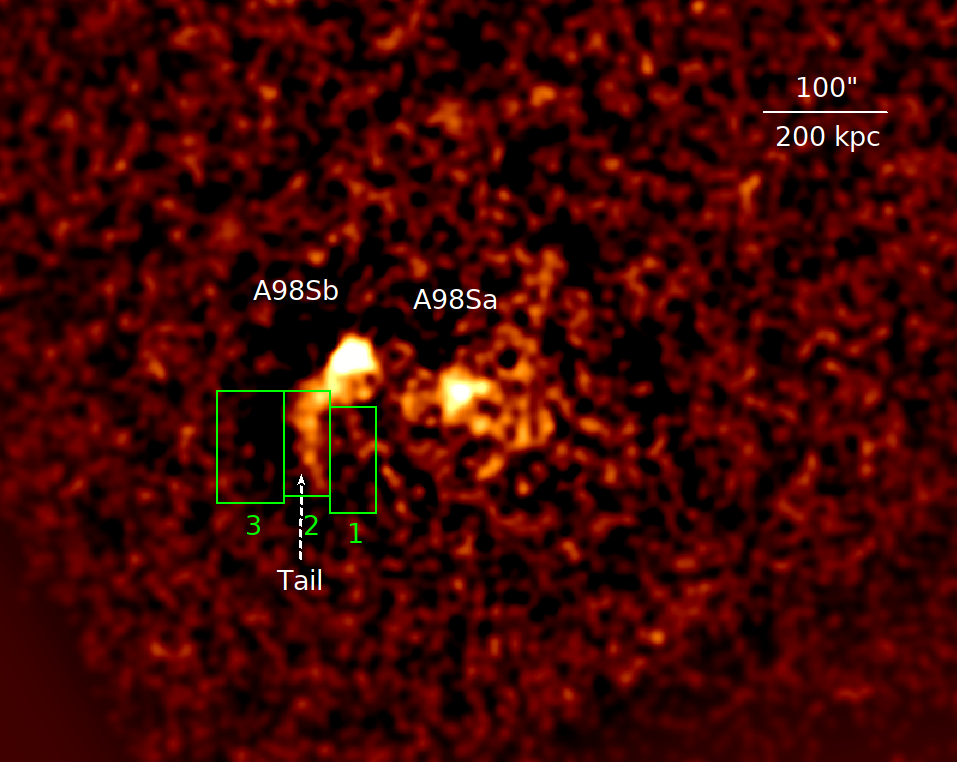} & 
         \includegraphics[scale=0.25]{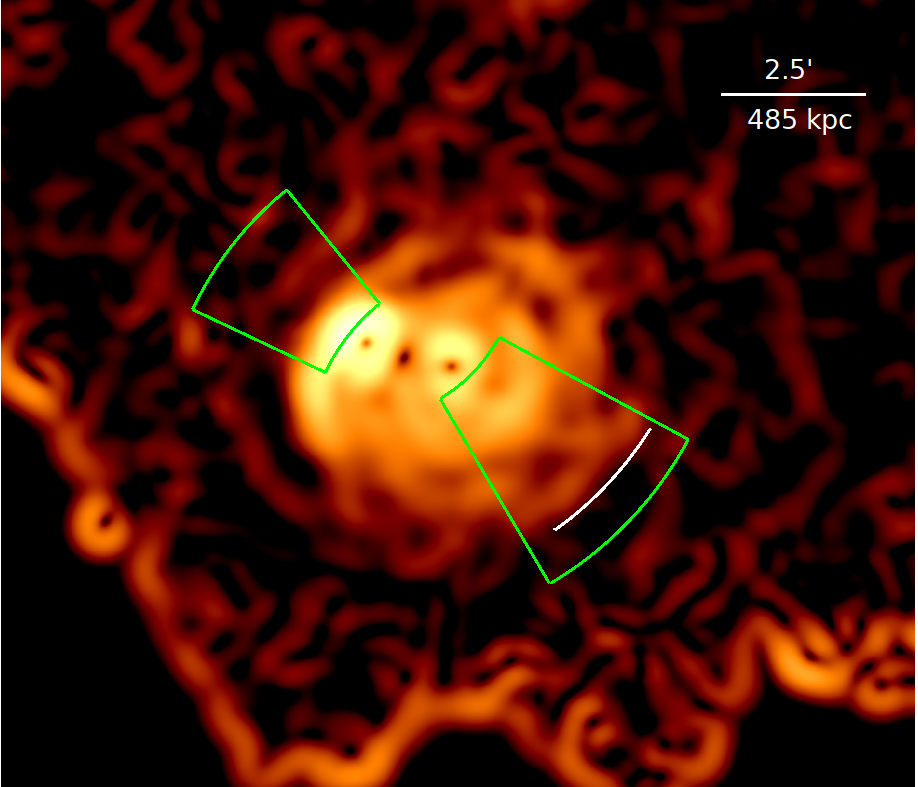}
\end{tabular}
\caption{{\it Left}: Unsharp-masked image of
A98Sa and A98Sb created by subtracting images smoothed
by 2D Gaussians with $\sigma$ = 2$\arcsec$ and 20$\arcsec$.
Green regions are used for the spectral analysis. The `tail'
is marked with white arrow.
{\it Right}: Gaussian gradient magnitude filtered image of A98S.
The image is smoothed by $\sigma$ = 13$\arcsec$. 
A sharp gradient is clearly visible to southwest of the A98Sa and
marked by a white curve. Green sectors are used for spectral analysis.}
\label{fig:A98S_unsharp_resid}
\end{figure*}

Figure \ref{fig:whle_img}
reveals that the central sub-cluster (A98S) 
hosts two distinct surface brightness peaks. 
\citet{2014ApJ...791..104P} 
showed that the western peak coincides with a 
WAT AGN.
Using the SDSS-r magnitude of the galaxy distribution of A98, 
they concluded that these two surface brightness 
peaks are likely to be associated 
with two BGCs from two merging subclusters.
We denote the western sub-cluster as A98Sa and 
the eastern sub-cluster as A98Sb. 
Figure \ref{fig:whle_img} 
exhibits that the X-ray emission 
is extended from east to west and appears to 
cut off abruptly northeast of the A98Sb. 
To better highlight these features, 
we create an unsharp-mask 
image in the following way. 
We smoothed the background subtracted 
and exposure corrected X-ray image 
by
2D Gaussian kernels with 
$\sigma$ = 2$\arcsec$ and 20$\arcsec$.
The images are then subtracted 
to obtain 
the unsharp-masked image, as shown 
in Figure \ref{fig:A98S_unsharp_resid} (left). 
The unsharp-masked image 
features the different substructures 
in the ICM of A98S. 
A98S appears to be separated into three 
concentrations of X-ray emitting gas: 
the two sub-cluster cores (A98Sa and A98Sb) 
and a ``tail" of emission to the southeast 
originating from the central region of A98Sb. 
The unsharp-masked image also reveals a 
surface brightness edge northeast of A98Sb. 
In contrast, A98Sa shows a relatively 
uniform morphology with no prominent 
surface brightness edges.

\begin{figure*}
\begin{tabular}{cc}
     \includegraphics[scale=0.4]{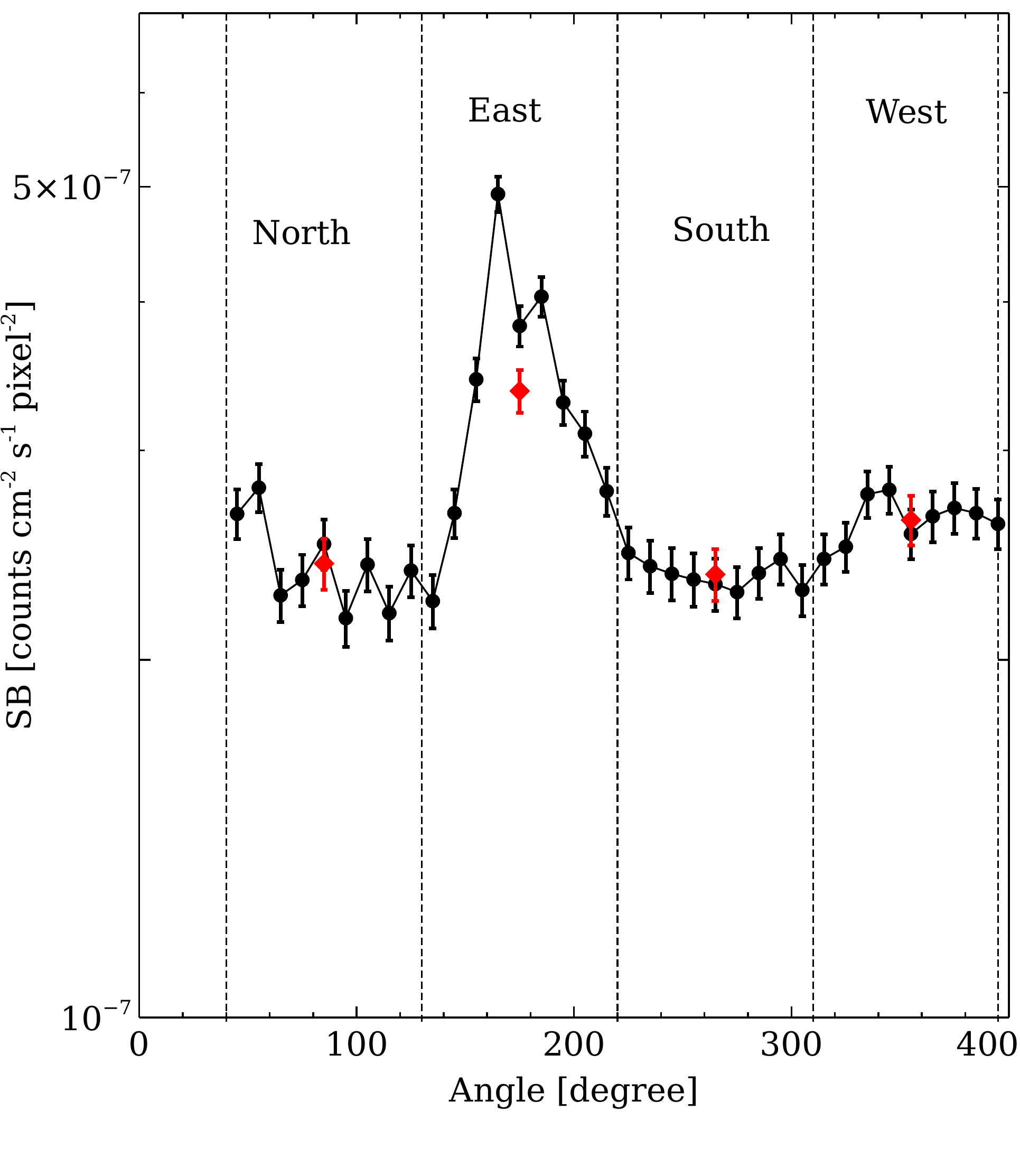} &
     \includegraphics[scale=0.33]{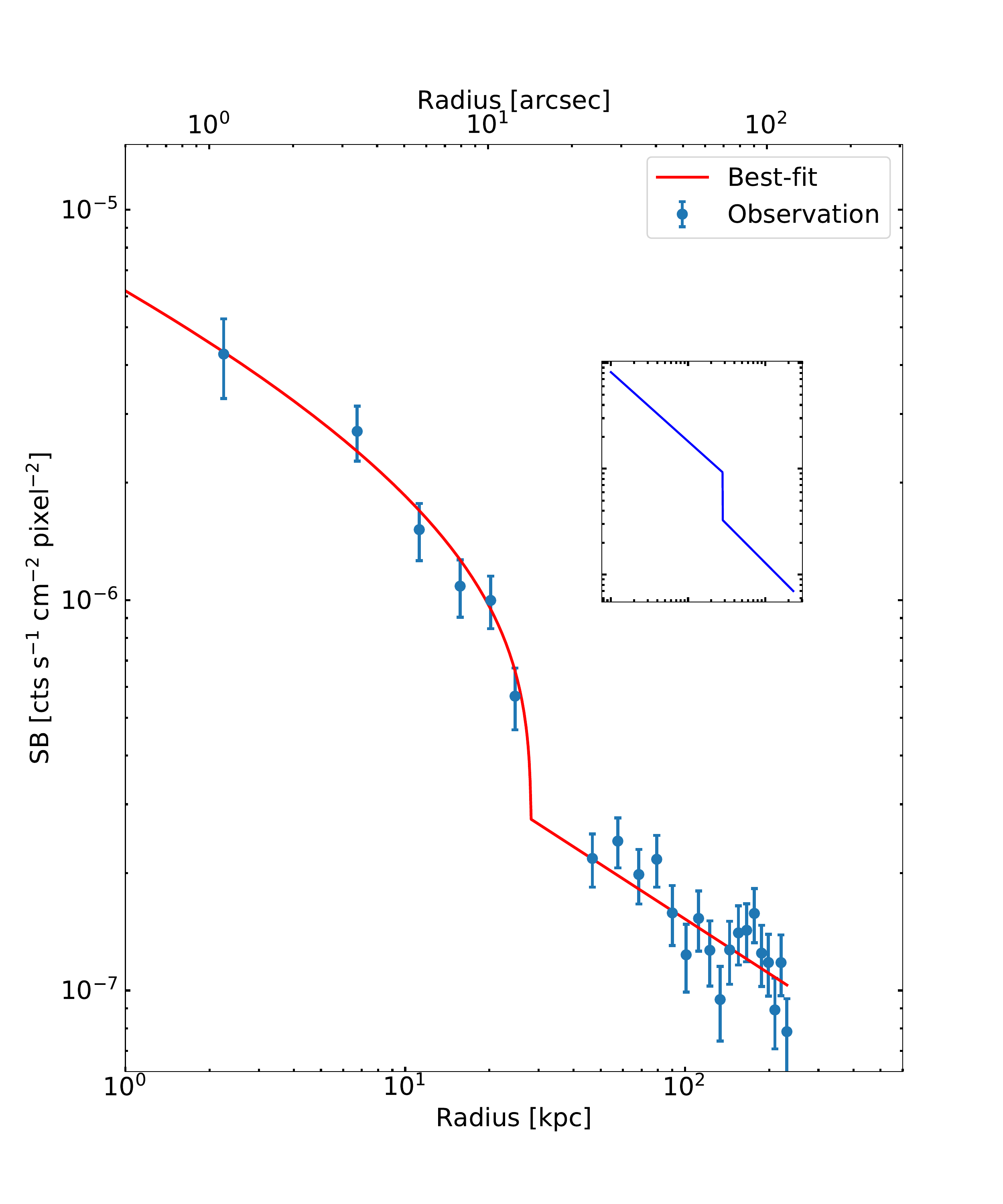}\\
\end{tabular}
\caption{{\it Left}: Azimuthal surface brightness profile 
in the 0.5 - 2 keV energy band centered on the A98Sa core. 
Black represents the surface brightness
at individual angles and red represents the binned surface
brightness in 90$^{\circ}$ sectors. The north and south sectors 
contain the northern and southern lobe of the WAT. 
East sector contains the second BCG, A98Sb.
{\it Right}:
Eastern sector (as shown in 
Figure \ref{fig:A98S_unsharp_resid}) 
surface brightness profile of A98Sb in the
0.5 - 2 keV energy band  fitted with a 
broken power-law model.  
Deprojected density profile is shown in 
inset figure.
}
\label{fig:a98s_sb}
\end{figure*}

We next examine the surface brightness edge 
seen in the northeast of A98Sb 
by extracting a surface brightness profile 
across the edge, 
as shown in Figure \ref{fig:a98s_sb} (right).
The surface brightness profile shows a 
discontinuity similar to what we have seen in A98N. 
We fit the profile with a broken power-law model, 
as detailed in Section \ref{sec:A98N}. 
Figure \ref{fig:a98s_sb} 
displays the best-fit model (in red) 
and 3D density profile (inset). 
We obtain a best-fit power-law index of 
$\alpha_1$ = 0.650 $\pm$ 0.003 and 
$\alpha_2$ = 0.7 $\pm$ 0.1. 
The associated density jump across the edge is 
$\rho_2$/$\rho_1$ = 2.8 $\pm$ 0.2
with a best-fit edge radius of 28 $\pm$ 2 kpc.

\begin{figure}
    \centering
    \includegraphics[scale=0.35]{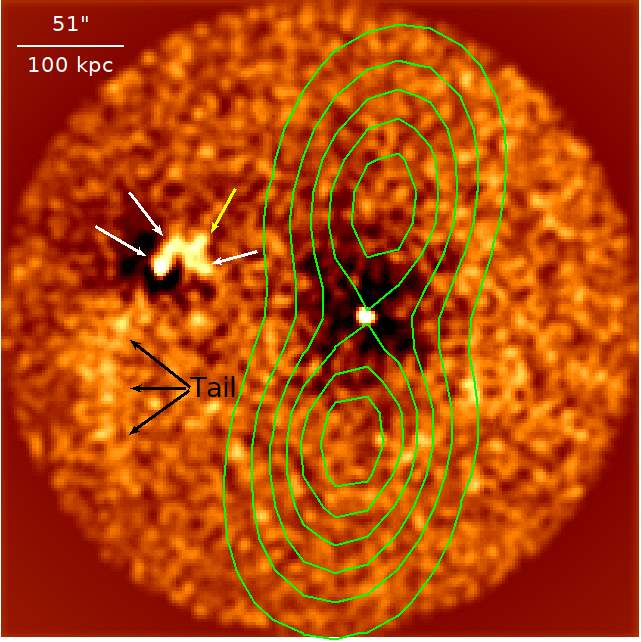}
    \caption{Residual image of excess X-ray 
    emission of A98Sa and A98Sb with 
    1.4 GHz radio contours
    overlaid in green. The image was created
    by subtracting two 2D elliptical $\beta$-models
    from an unbinned image with point sources excluded.
    White, yellow, black arrows indicate the spiral structure,
    the bifurcation of spiral pattern, and 
    the `tail' of X-ray emission, respectively.}
    \label{fig:resid_a98s}
\end{figure}

To further investigate
any faint substructure associated
with both sub-cluster cores (A98Sa and A98Sb), 
we obtain a $\beta$-model subtracted image of A98S. 
We use two 2D elliptical $\beta$-models centered 
at both sub-cluster cores to fit the surface 
brightness image in the 0.5-2 keV energy band. 
We also include a constant in the model 
to account for the sky background 
and any residual particle background.
The model is fitted in {\tt Sherpa}, 
and the best-fit parameters are 
obtained by minimizing the Cash statistic
\citep{1979ApJ...228..939C}. 
The best-fit 2D $\beta$-model image is then 
subtracted from the source image of A98S, 
leaving the residual image 
shown in Figure \ref{fig:resid_a98s}.

Figure \ref{fig:resid_a98s}
reveals a spiral structure likely to 
be associated with the gas sloshing resulting 
from a recent fly-by of a subhalo 
near the central region of A98Sb. 
The excess emission from the ``tail" seen in 
the residual and unsharp-masked image
may suggest a trail of materials that 
has been ram pressure stripped in the 
gravitational potential. 
Alternatively,
the spiral structure and the tail
may be due to ram pressure stripping, 
as the subcluster orbits in the main 
cluster's potential, 
and the tail is potentially influenced 
by ICM ``weather" (e.g. bulk turbulence).
Either scenario could be true. 
A further deeper observation is
required to probe the feature in
detail.

Compared to A98Sb, A98Sa shows a 
relatively regular 
morphology. 
Since there are clear radio lobes, 
one might expect cavities in the ICM 
evacuated by these lobes 
\citep{2008ApJ...685..128M}.  
A weak detection of cavities at the 
location of the lobes was reported in
\citet{2014ApJ...791..104P}.
We therefore extract an azimuthal surface brightness 
profile, shown in Figure \ref{fig:a98s_sb},
centering at the A98Sa core with 
a radius of $\sim$ 150$\arcsec$. 
The sharp increase in the surface brightness 
to the east with a peak at $\sim$ 165$^{\circ}$ 
is due to the surface brightness 
peak associated with A98Sb. 
Similar results were obtained by 
\citet{2014ApJ...791..104P}. 
However, we do not find any significant 
difference between the surface brightness 
in any other directions. 
Since A98Sa hosts a WAT radio source, 
the absence of any substantial decrement 
in the surface brightness at north and 
south directions indicates that either 
the radio lobes are filled with X-ray emitting gas, 
or the jet axis is far enough from 
the plan of the sky to prevent a detection.
Several previous studies also found jets 
and lobes in other merging galaxy clusters, 
such as 
A2199 \citep{2013ApJ...775..117N}, 
A1682 \citep{2019A&A...627A.176C}, 
A1446 \citep{2008ApJ...673..763D}, 
A562 \citep{2011ApJ...743..199D,2020AJ....160..152G}, 
and A1775 \citep{2021ApJ...913....8H}. 

The GGM image of A98S shown in Figure 
\ref{fig:A98S_unsharp_resid} 
features a
possible surface brightness edge at about 370 kpc
southwest of A98Sa. We extract a surface 
brightness profile across the edge and fit that profile
using a broken power-law model, as seen in Figure
\ref{fig:A98sa_west_temp}. We obtain best-fit 
power-law indices of 
$\alpha_1$ = 0.7 $\pm$ 0.1 and 
$\alpha_2$ = 0.9 $\pm$ 0.2.
The electron density jumps by a factor of
1.7 $\pm$ 0.5 across the edge. We find the 
best-fit edge radius of 375 $\pm$ 10 kpc 
coincides with the apparent rapid 
change in the surface brightness gradient
seen in GGM image (Figure \ref{fig:A98S_unsharp_resid}).
We discuss the spectral properties 
of the surface brightness edges seen in A98Sa and A98Sb
in Section 
\ref{sec:A98S_spectral}.

\section{Spectral analysis} \label{sec:spectral}
\subsection{Global ICM properties}
To constrain the global properties of both sub-clusters, 
we first extracted spectra from two circular regions 
centering on each sub-cluster core with a radius of $\sim$ 
2$'$ for A98N and 3$'$ for A98S
(roughly similar regions as seen in \citealt{2014ApJ...791..104P}). 
Both regions contain $\gtrsim$ 32000 background-subtracted 
counts in the 0.6-7 keV energy band and contain a vast 
majority of the cluster emission after eliminating the point sources, 
which is imperative to probe the global properties of both sub-clusters. 
The background spectra were hard-band scaled (10--12 keV) and 
subtracted from the source spectra before fitting. 
We restricted the spectral fitting to the 0.6-7 keV energy band and 
binned the spectra to contain a minimum of 40 counts per bin. 
We fixed the redshift to z = 0.1042. 
The spectral analysis was performed using {\tt Xspec} version 12.11.1. 
All resolved point sources were excluded before the spectral analysis. 
We used solar abundance table given in \citet{2009ARA&A..47..481A}.
For each sub-cluster, a single-temperature fit to the observed spectra 
using an absorbed thermal plasma emission model 
(i.e. {\tt phabs}$\times${\tt apec}) produced a good fit \citep{2001ApJ...556L..91S}.

For A98N, we measured a temperature of 3.05 $\pm$ 0.09 keV 
and a metal abundance of 
0.63 $\pm$ 0.10 Z$_{\odot}$ with a 
$\chi^2$/dof of 786/761. 
Additionally, spectral fitting to the 
central sub-cluster (A98S) yielded a 
best-fit temperature of 3.04 $\pm$ 0.12 keV
and 
an abundance of 0.42 $\pm$ 0.10 Z$_{\odot}$ 
with a $\chi^2$/dof of 870/886. 
In both cases, the 
hydrogen column density was fixed to 
the Galactic value, N$_{H}$ = 3.06$\times$10$^{20}$ cm$^{-2}$, 
estimated using the LAB survey \citep{2005A&A...440..775K}. 
\citet{2014ApJ...791..104P} also reported the temperature and abundance of both 
sub-clusters using a 38 ks $\chandra$ observation and a 37 ks $\xmm$ observation. 
For both sub-clusters, our results better
determine the gas temperatures 
and abundances, but are consistent with 
the earlier measurement by
\citet{2014ApJ...791..104P}. 

\subsection{Temperature map}
For understanding the thermodynamic structures 
of the ICM in 
A98N and A98S, we constructed a temperature map following 
the techniques described in 
\citet{2008ApJ...688..208R,2009ApJ...696.1431R}. 
We 
extracted an elliptical region
containing both 
sub-clusters and binned the image with 
10 background-subtracted counts per 
pixel to aid the computational speed. 
For each pixel, we extracted spectra from a circular 
region with a radius set such that the region contains 
$\sim$ 2300 background-subtracted source counts in the 
0.6-7 keV energy band. 
This particular number of 
source counts 
was adopted to ensure that the uncertainty in 
temperature measurements in the fainter regions of the ICM 
was 
$\lesssim$ 25\%. 
We generated appropriate response and ancillary 
response files for each region, and the background spectra 
were subtracted from the total spectra. 

\begin{figure}
    \centering
    \includegraphics[scale=0.37]{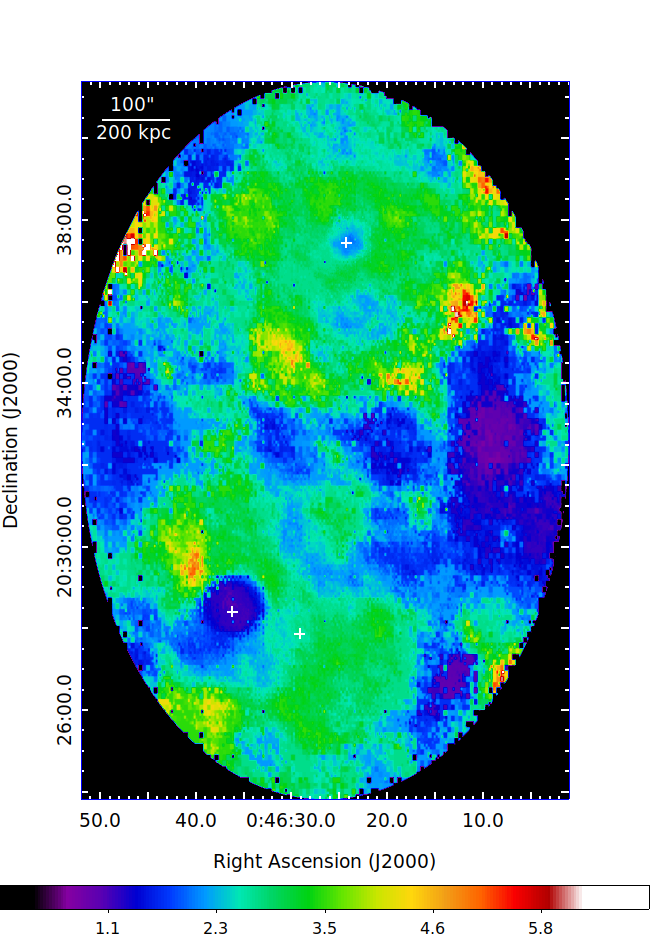}
    \caption{Projected temperature map of A98N/A98S.
    Units are in keV. The individual sub-cluster centers
    are marked by white crosses. A98N is at RA = 0:46:25.0,
    Dec = 20:38:00, A98Sa is at RA = 0:46:35.0, Dec = 20:28:03,
    and A98Sb is at RA = 0:46:28.0, Dec = 20:28:00.
    The warmer gas `arc' can be seen
    to the south of A98N. A98N and A98Sb have cool central cores, 
    whereas A98Sa has a warmer, roughly isothermal core.}
    \label{fig:temp_map}
\end{figure}

Figure \ref{fig:temp_map} shows the temperature map of A98N/A98S. 
A98N and A98Sb 
appear to have cool cores with hot gas enclosing them, 
which is expected for cool core clusters. 
In contrast, A98Sa appears to have a warm core with 
a uniform gas temperature of $\sim$ 
3.7 keV within 20 kpc of the core. 
This suggests that A98a was 
transformed into a non cool-core cluster 
during an earlier subcluster merger.  

The northern sub-cluster has an asymmetric `arc' of 
hot gas with a temperature of $>$ 3.5 keV,
which is clearly visible in Figure \ref{fig:temp_map}. 
This arc radius ($\sim$ 400 kpc)
is consistent with 
the radius of the shock front
that we reported in paper I \citep{2022ApJ...935L..23S},
suggesting the `arc' consists of shock heated gas. 
The uncertainties in the temperature 
map range from $\lesssim$ 15\% for brighter regions 
and are between 
20-25\% for the fainter regions. Since each region in 
the temperature map contains a fixed source count, 
the extraction regions 
are larger in the fainter parts of the ICM, 
making the temperature map highly smoothed. 


\subsection{A98N: East and North cold fronts}
We extracted spectra from the 
eastern and northern sectors in A98N 
to measure the temperature and density changes across the 
surface brightness edges seen in Figure \ref{fig:edge_fitting}. 
Each sector was divided into five individual bins. 
These bins were positioned to measure the gas 
properties on either side of the surface brightness edges 
while maintaining a minimum of 3000 source 
counts in each extracted spectrum. 
We set this lower limit to guarantee enough photon 
counts for good spectral fitting and constraints 
on the parameters. 
Spectra were extracted from those bins in each sector.

\begin{figure*}
\gridline{\fig{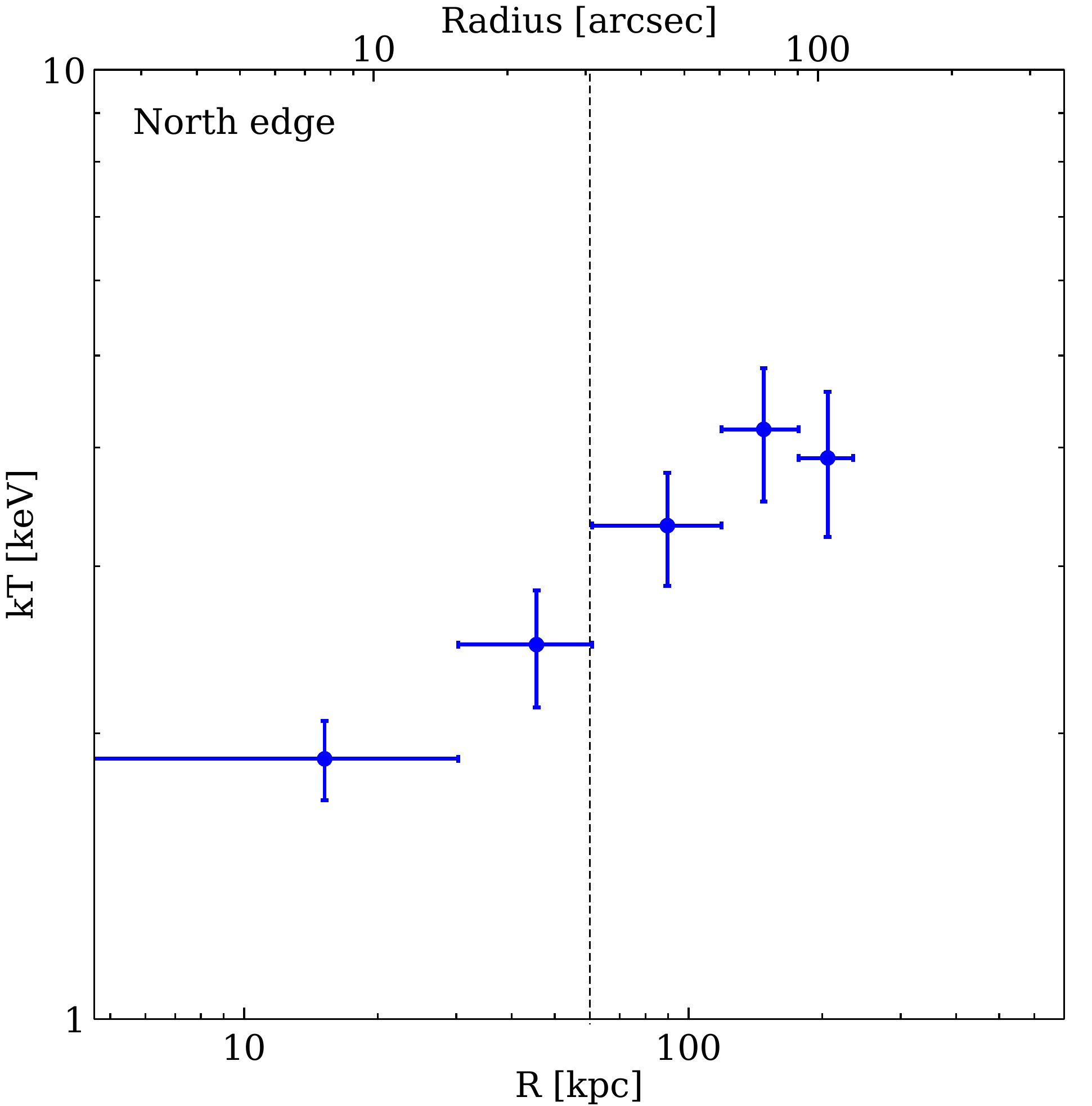}{0.33\textwidth}{(a) Temperature}
          \fig{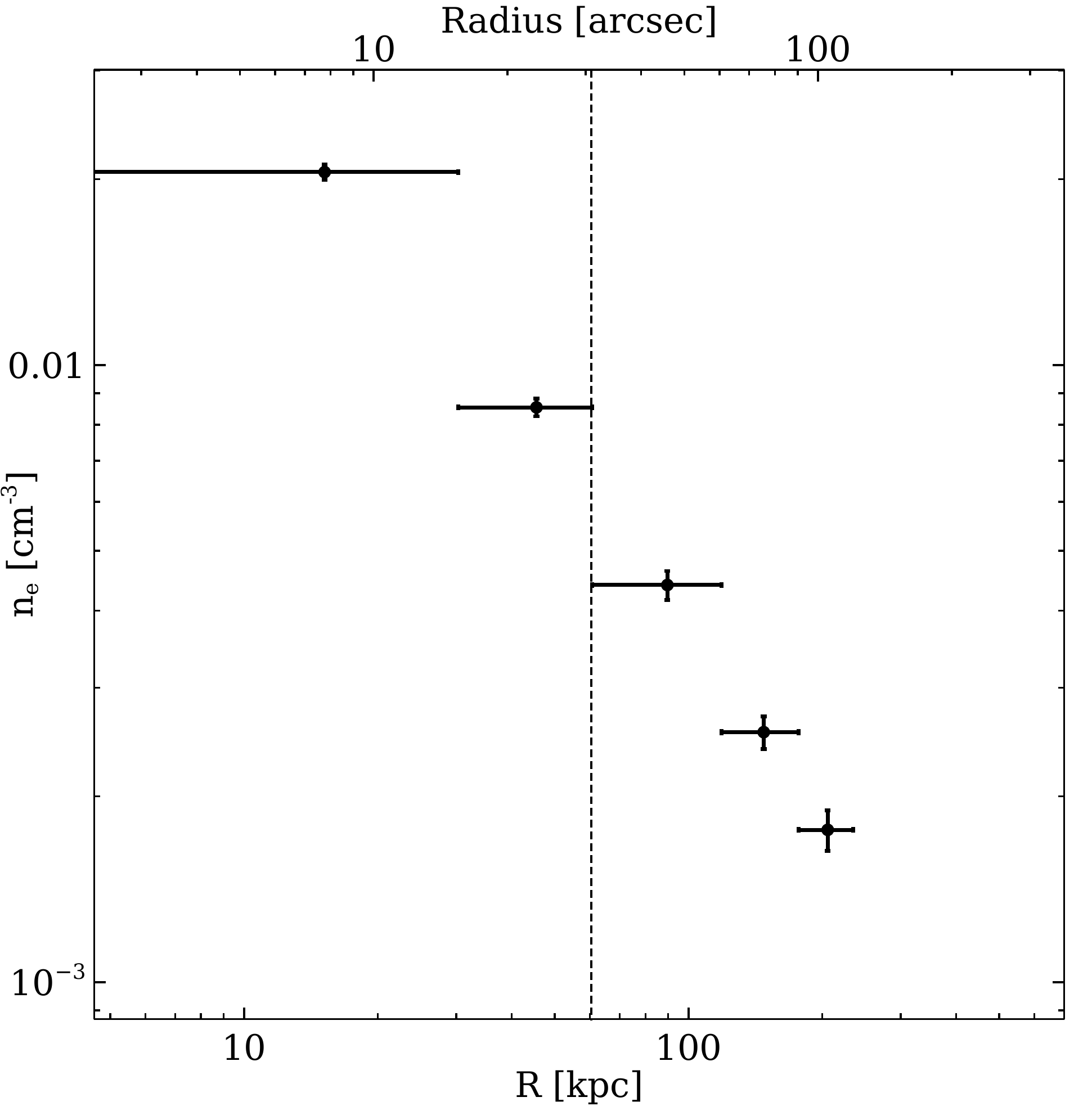}{0.33\textwidth}{Density}
          \fig{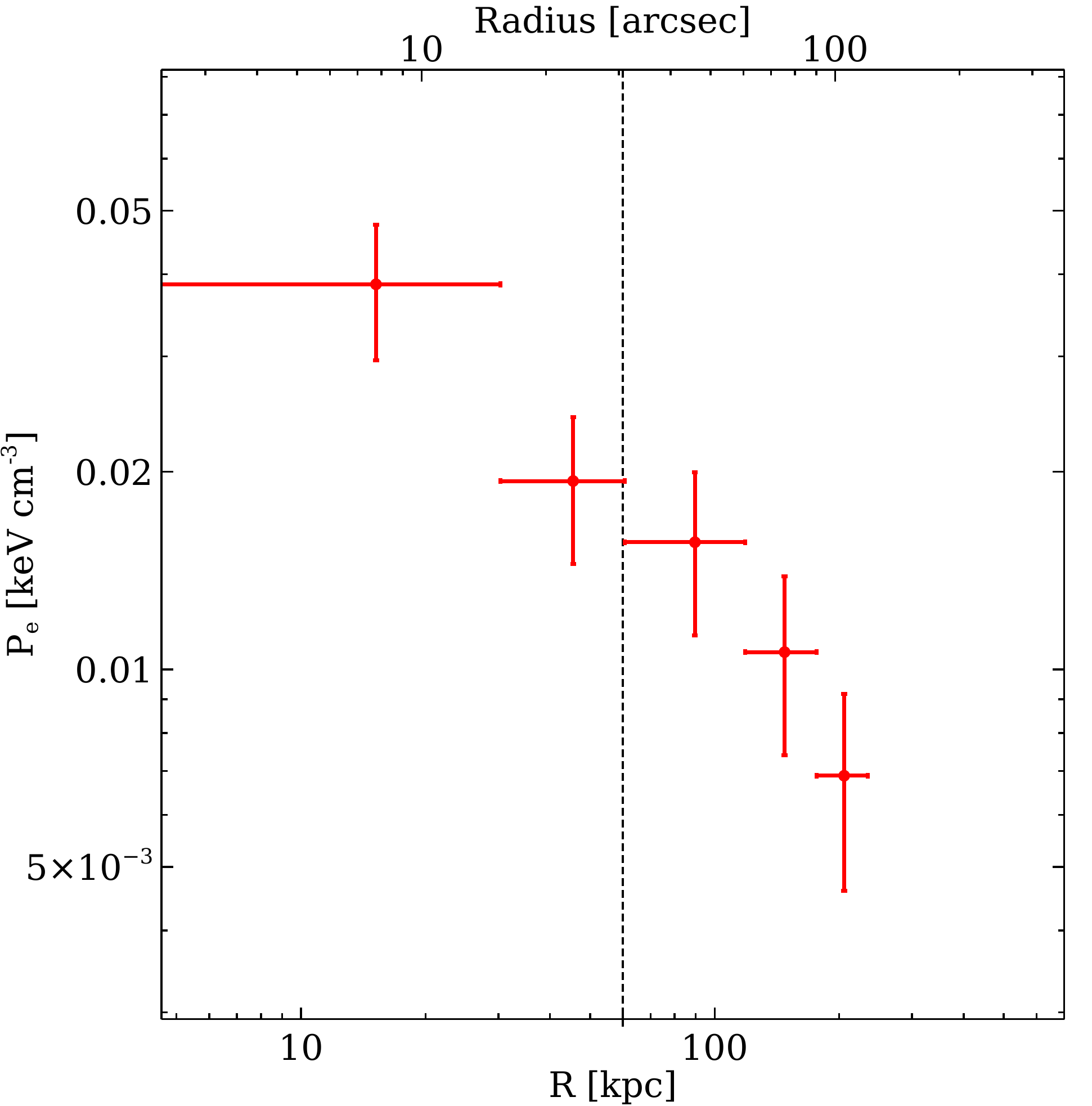}{0.33\textwidth}{Pressure}
         }
\gridline{\fig{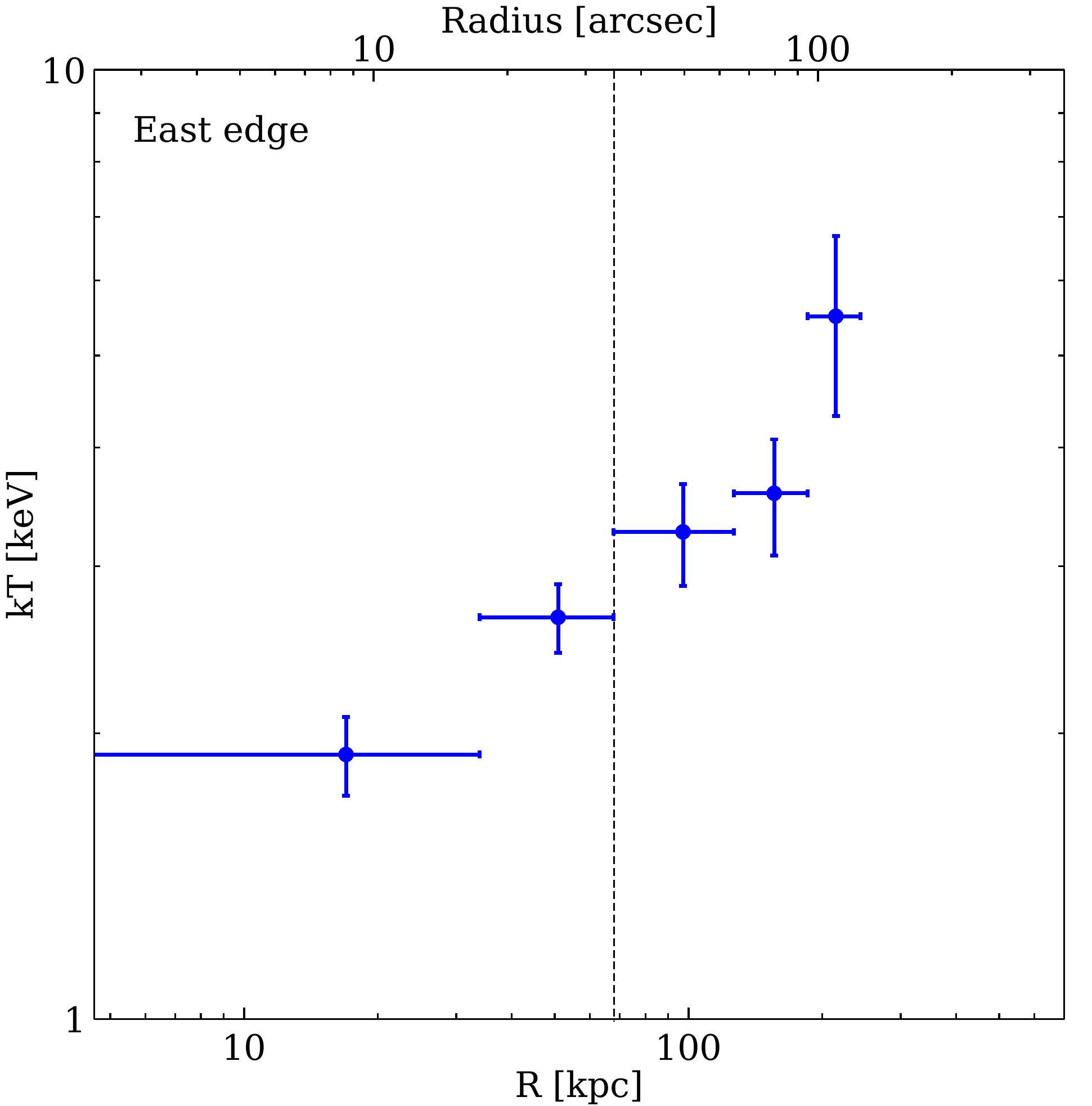}{0.33\textwidth}{(b) Temperature}
          \fig{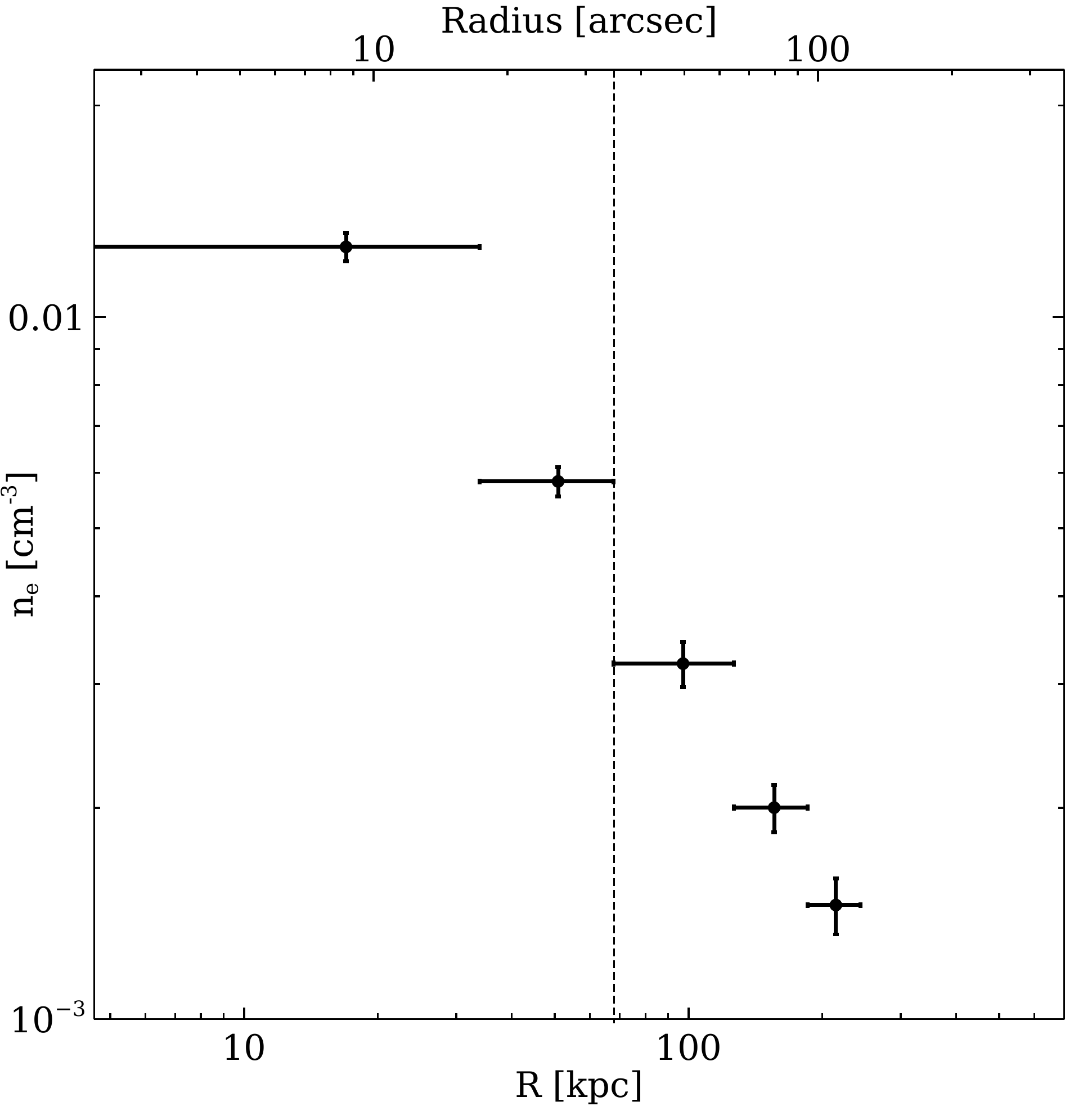}{0.33\textwidth}{Density}
          \fig{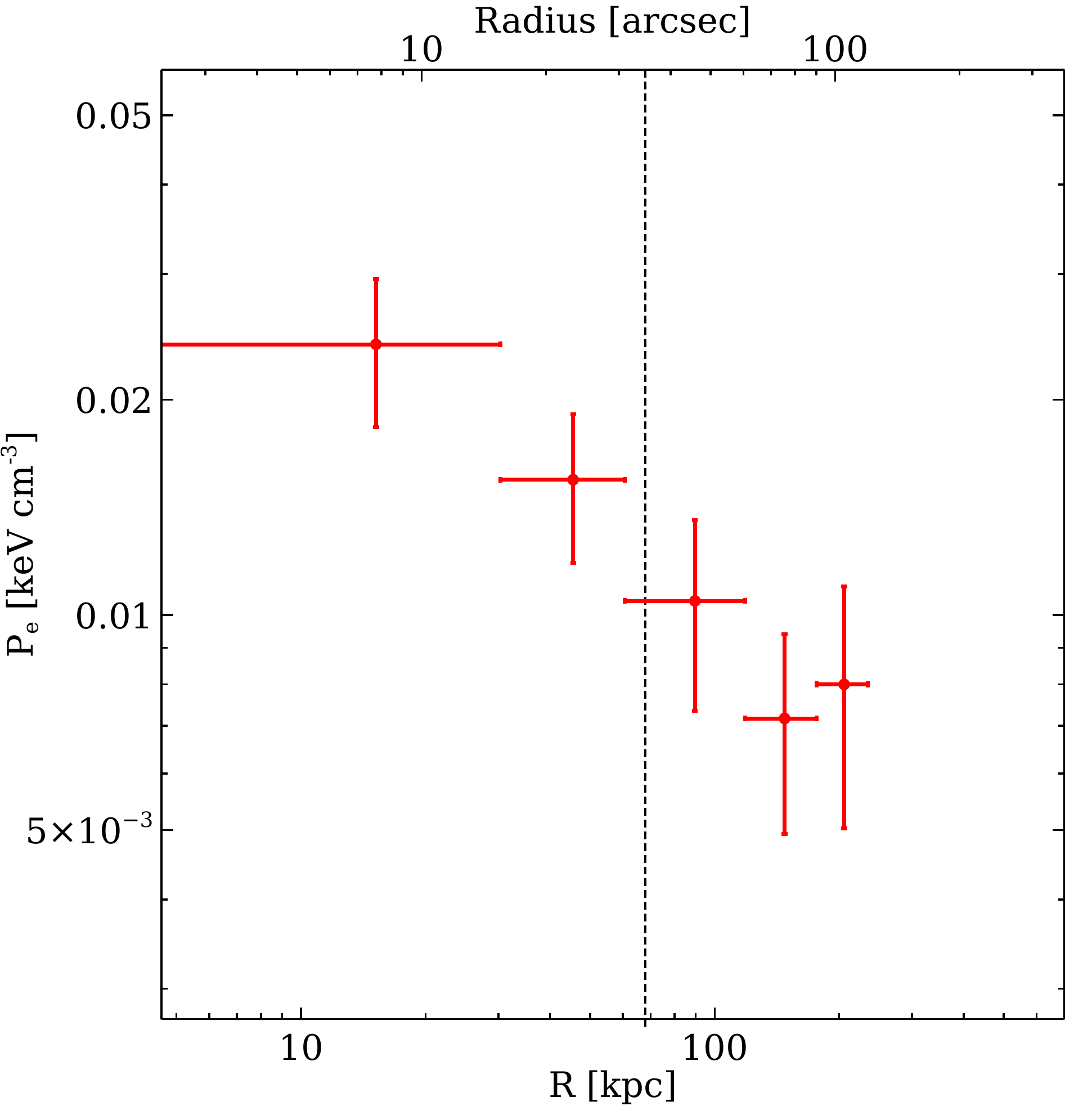}{0.33\textwidth}{Pressure}
          }
\caption{ {\it Top panel}: Northern sector projected temperature,
deprojected electron density, and projected electron pressure
profiles centered on the A98N core. The metallicity was fixed to
0.4 Z$_{\odot}$. The projected pressure profile was estimated
using the deprojected densities and projected temperatures.
The vertical dashed line represents the position of the surface brightness
edges. {\it Bottom panel}: similar as above but for eastern sector.
\label{fig:edge_temp}}
\end{figure*}

We fixed the metallicity to an 
average value of 0.4 Z$_{\odot}$ since it was poorly 
constrained if left free \citep{2010MNRAS.406.1721R}. 
The best-fit parameters were obtained by minimizing
the Cash statistics. 
Figure \ref{fig:edge_temp} 
shows the radial profiles of the 
best-fit projected temperatures 
and pressures
obtained 
across the eastern and northern edges. 
The temperature of the gas increases from 
the A98N core to the larger radii in both directions. 
Our measurements confirm the 
electron density jumps detected in the surface 
brightness profile analysis, 
but the temperature and pressure 
profiles appear continuous across the edges.  
However, the cooler gas inside of the edges is more consistent 
with a cold front interpretation.
Similar cold fronts are also found in other 
clusters such as the 
Bullet cluster \citep{2002ApJ...567L..27M}, 
A2142 \citep{2000ApJ...541..542M}, 
A3667 \citep{2001ApJ...551..160V}, 
A2146 \citep{2010MNRAS.406.1721R}, 
A2256 \citep{2020MNRAS.497.4704G}, 
A2554 \citep{2019MNRAS.488.2917E}, 
Perseus \citep{2017MNRAS.468.2506W,2020arXiv200614043W},
and in many others \citep{2018MNRAS.476.5591B}.

\subsection{A98S: cold front, tail, and western SB edge} \label{sec:A98S_spectral}
A98S consists of two individual sub-clusters, A98Sa and A98Sb. 
The temperature map in Figure \ref{fig:temp_map} 
shows 
that the central region of A98Sb 
is cooler than A98Sa. 
Previous $\chandra$ observations by 
\citet{2014ApJ...791..104P}
showed that A98Sb is cooler than A98Sa at a 
3-$\sigma$ level. 
With 
our deeper $\chandra$ observations,
we examined the spectra of each sub-cluster
to measure the 
global ICM temperatures of individual sub-clusters. 
We extracted spectra 
from a circular region with 50$\arcsec$ radius and 
centering at both sub-clusters cores. 
These regions were chosen based on the peaks
of the X-ray emission, 
assuming each of them includes the prominent extended 
emission from the 
corresponding sub-cluster. 
Point sources and background spectra were 
analyzed as described in Section \ref{sec:spectral}.
We measured a temperature of 3.70$_{-0.28}^{+0.33}$ keV for 
the western sub-cluster (A98Sa) and 
2.47$_{-0.20}^{0.21}$ for the eastern 
sub-cluster (A98Sb). 
Our measured temperatures are consistent 
with that of \citet{2014ApJ...791..104P}
within their uncertainties 
(90\% confidence limit). 
We also found that the A98Sb 
is cooler than the A98Sa at a 5-$\sigma$ level. 
This implies that A98Sb has a remnant cool core, 
whereas the core of A98Sa has been disrupted, 
so that this is a non cool-core cluster. 
Since A98Sa hosts a WAT radio source 
\citep[4C 20.04;][]{1993ApJ...408..428O}, 
the surrounding ICM may be heated by AGN jets. 
Similar ICM heating through AGN feedback 
is also found in other galaxy clusters 
such as A2052
\citep{2011ApJ...737...99B}, Perseus
\citep{2003MNRAS.344L..43F}, A2626
\citep{2018A&A...610A..89I,2020A&A...643A.172I}, 
and A1795
\citep{2018A&A...618A.152K}.
Alternatively, the core of
A98Sa    
may have been disrupted by either the 
ongoing merger or a previous merger.

We next examined the ``tail" like substructure 
of the eastern sub-cluster that appeared in 
the unsharp masked image (Figure \ref{fig:A98S_unsharp_resid}). 
We extracted spectra from three individual regions, 
as shown in Figure \ref{fig:A98S_unsharp_resid}.
For region 2, 
this yielded a temperature of 
$kT$ = 1.8$^{+0.3}_{-0.2}$ keV and 
an abundance of $Z$ = 0.35$^{+0.49}_{-0.24}$ Z$_{\odot}$. 
Due to low photon counts in 
regions 1 and 3, we linked the parameters for 
those two regions while fitting, 
assuming they have similar temperature and 
abundance. 
For regions 1 and 3 together, we measured a 
temperature of $kT$ = 3.1$^{+0.6}_{-0.4}$ keV 
and an abundance of $Z$ = 0.80$^{+1.01}_{-0.57}$ Z$_{\odot}$. 
We found the tail is significantly 
cooler than the surrounding gas 
at about 4.2-$\sigma$ level, 
which suggests that it is likely to be a cool-core 
remnant that is being ram pressure stripped. 

\begin{figure*}
\gridline{\fig{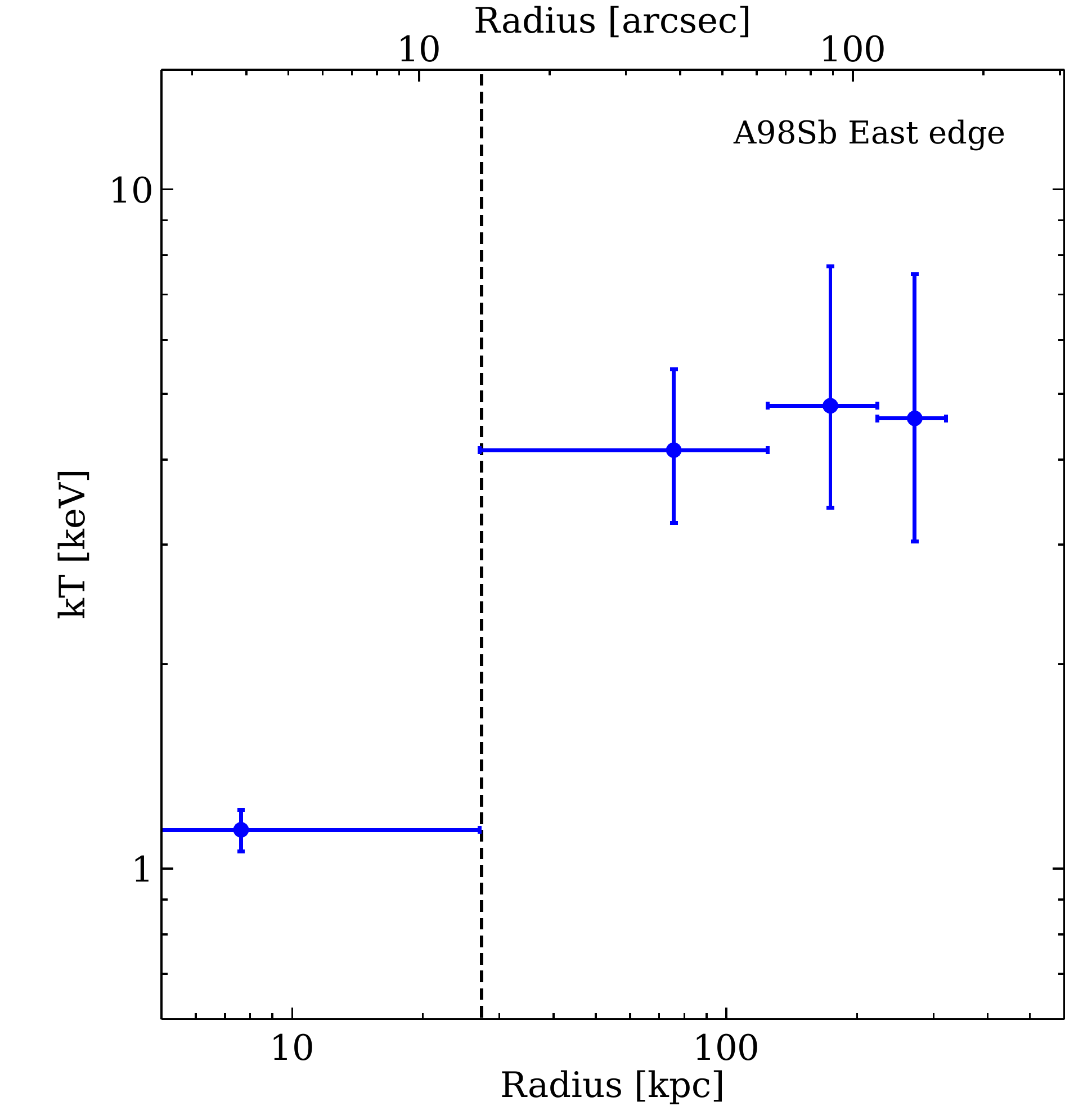}{0.33\textwidth}{Temperature}
          \fig{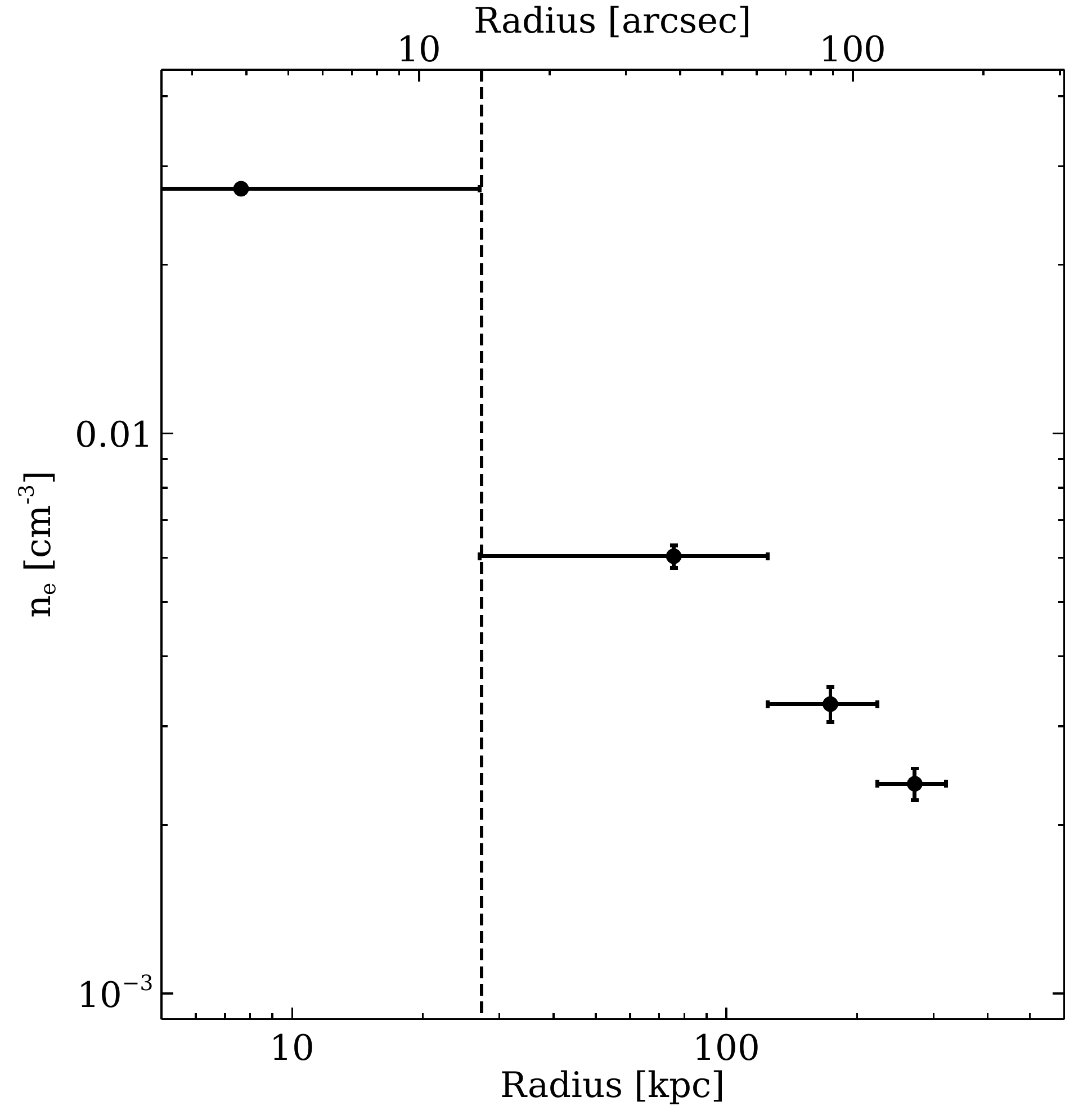}{0.33\textwidth}{Density}
          \fig{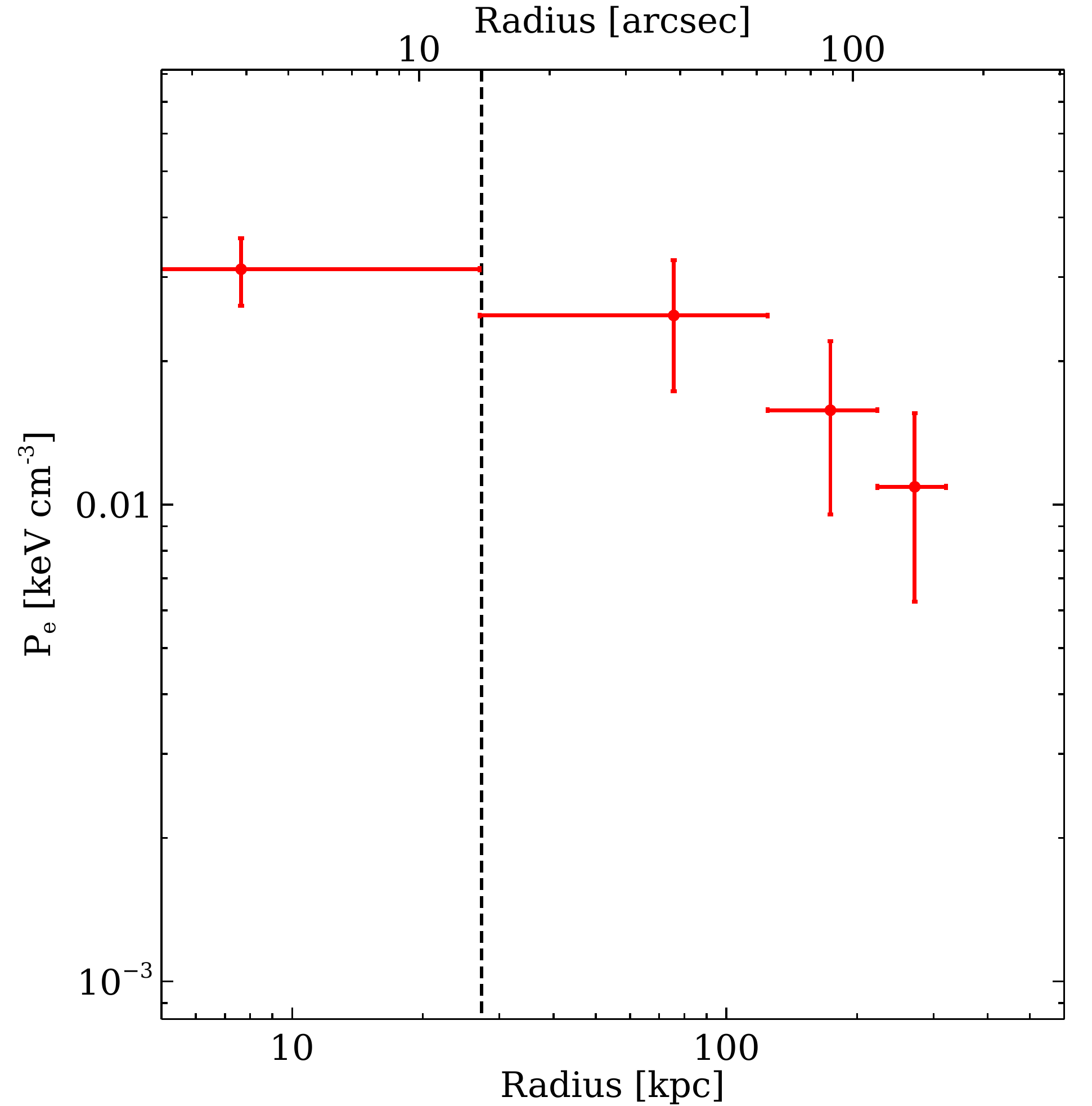}{0.33\textwidth}{Pressure}
         }
    \caption{ Eastern sector projected temperature,
deprojected electron density, and projected electron pressure
profiles centered on the A98Sb core. The metallicity was fixed to
0.4 Z$_{\odot}$. The projected pressure profile was estimated
using the deprojected densities and projected temperatures.
Vertical dashed line represents the position of the surface brightness
edge.
    }
    \label{fig:A98Sb_fitt_temp}
\end{figure*}

\begin{figure*}
\begin{tabular}{cc}
     \includegraphics[scale=0.33]{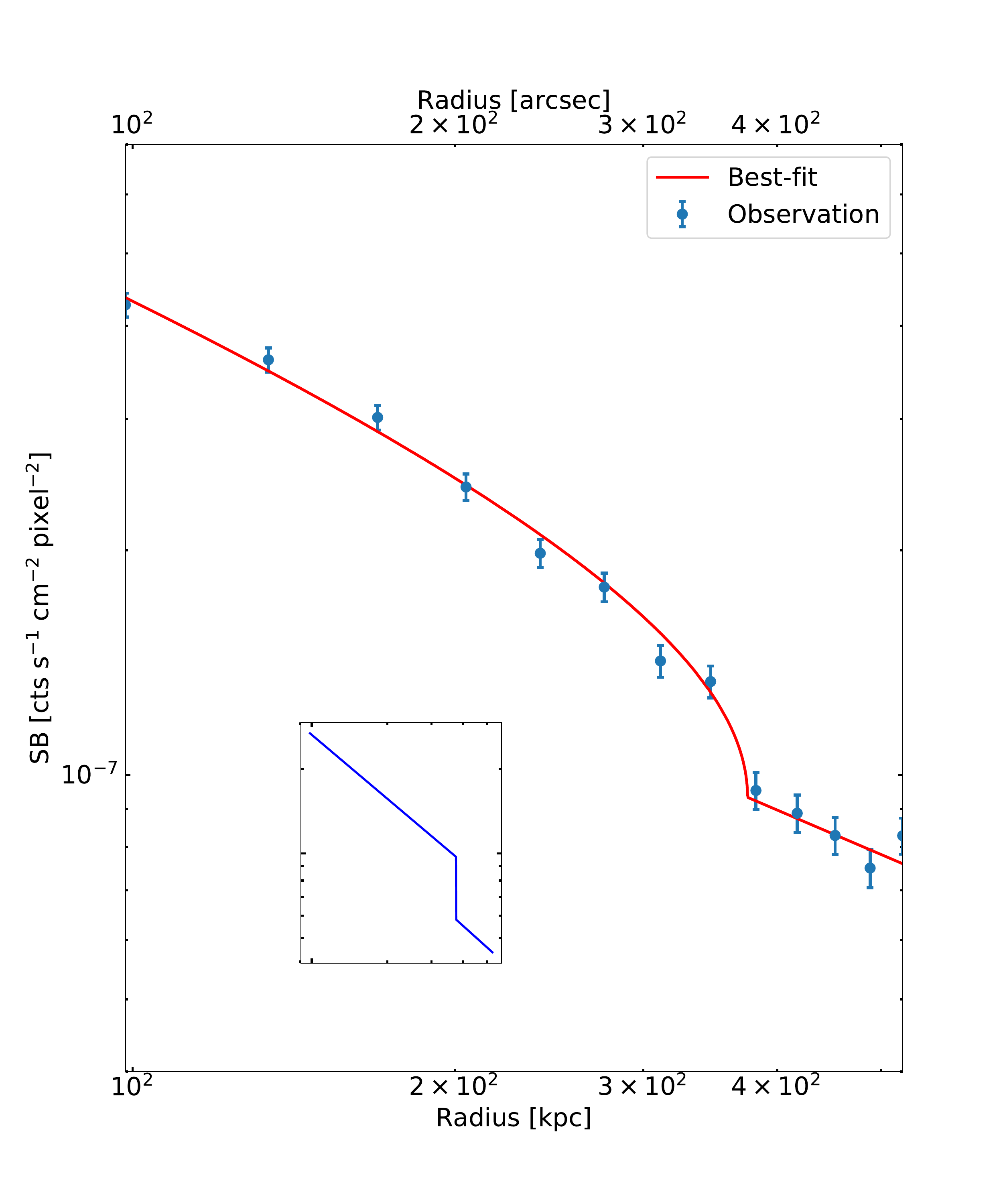} &
     \hspace{-20pt}
     \includegraphics[scale=0.43]{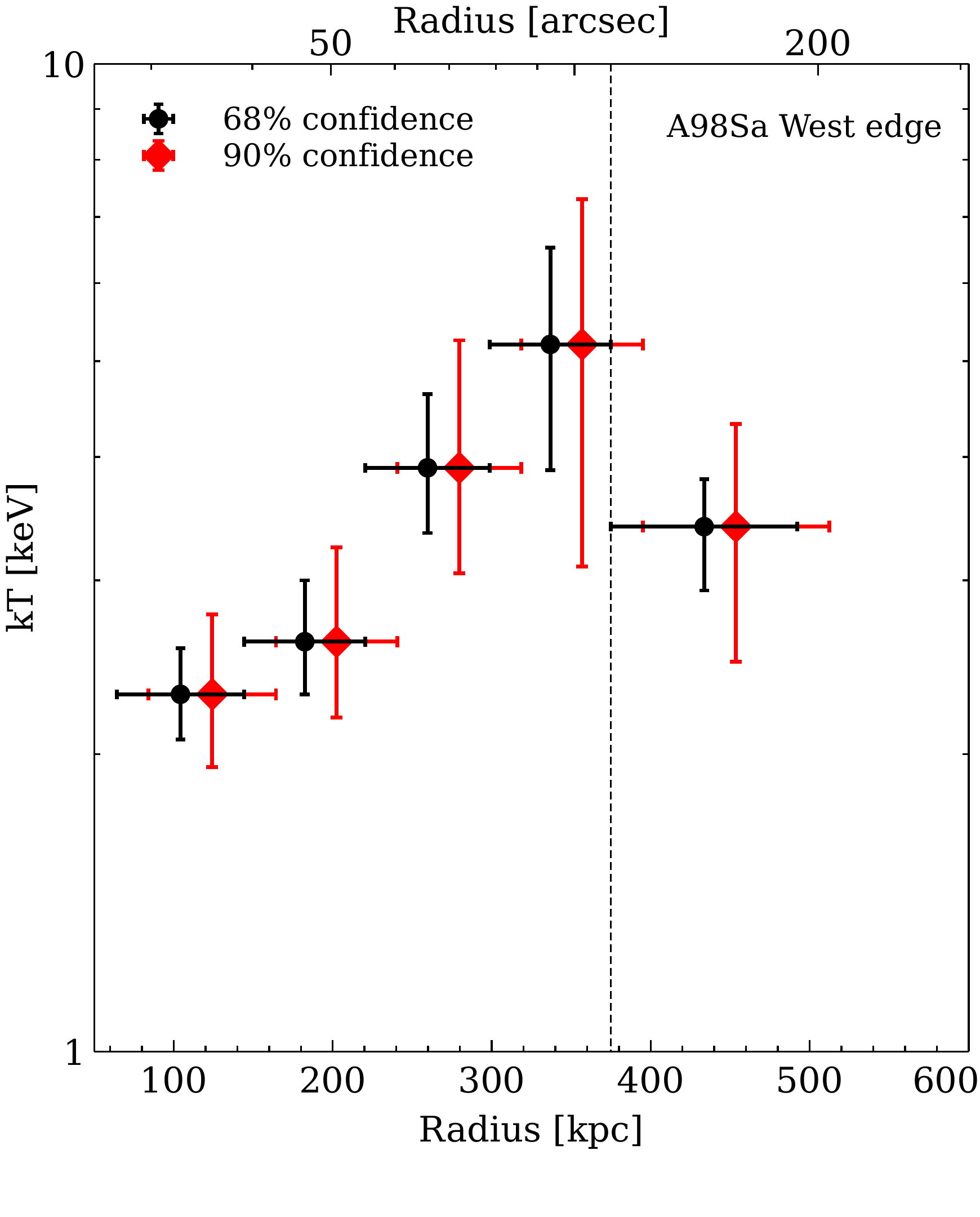}\\

\end{tabular}
    \caption{{\it Left}: western sector surface brightness profile of A98Sa
    in the 
    0.5-2 keV energy band fitted with a broken power-law model. 
    Deprojected density profile is shown in inset figure.
    {\it Right}:
    western sector projected temperature profiles
    of A98Sa. Black and red represent the uncertainties 
    estimated with 68\% and 90\% confidence level,
    respectively.
    The red data points are shifted to the right 
    for the purpose of clarity and comparison.
    Vertical dashed line represents
    the position of the surface brightness edge.}
    \label{fig:A98sa_west_temp}
\end{figure*}

Figure \ref{fig:a98s_sb} 
shows that A98Sb has a surface brightness edge 
in the northeast direction at about 28 kpc from its core. 
To measure the temperature across the surface brightness edge, 
we divided the eastern sector into four individual 
regions and extracted spectra from them. 
Those regions were drawn to contain a minimum of 
1000 background-subtracted photon counts in each 
extracted spectrum. 
We set this lower limit to ensure reasonable 
constraints on the parameters. 
We fitted each spectrum with
the abundance fixed to $Z$ = 0.4 Z$_{\odot}$. 
Figure \ref{fig:A98Sb_fitt_temp} 
shows the best-fit projected temperature 
profile of the eastern sector. 
The projected temperature of the gas increases rapidly 
at about 28 kpc, by a factor of $\sim$ 3.6. 
This temperature jump coincides with the surface brightness 
edge at 28 kpc, as shown in Figure 
\ref{fig:A98Sb_fitt_temp}, 
and corresponds to a drop in the 
deprojected electron 
density by a factor of $\sim$ 2.8. 
Figure \ref{fig:A98Sb_fitt_temp}
also shows the projected 
electron pressure profile 
of the eastern sector. 
We found that the electron pressure is roughly 
constant across the surface brightness edge.  
We, therefore, confirm this edge as a cold 
front contact discontinuity.

We next examined the surface brightness edge 
shown in Figure \ref{fig:A98sa_west_temp} (left)
southwest of A98Sa by extracting spectra from 
that sector dividing it into five 
individual regions. 
We fitted the spectrum from each region 
with the abundance fixed to 0.4 Z$_{\odot}$.
The best-fit projected temperature profiles 
are shown in Figure \ref{fig:A98sa_west_temp} (right)
with different levels of uncertainty. 
Considering both levels of uncertainties,
we found that the temperature profile rises 
to $kT$ = 5.1 keV at about $\sim$ 375 kpc, then drops suddenly
to $kT$ = 3.3 keV with a jump factor of $\sim$ 1.5.  
Taken all together, this is consistent with 
a shock. 
From the 3D density jump factor we obtained
a Mach number of the shock to be
$\cal{M}$ $\approx$ 1.5, consistent
with the Mach number estimated from the
temperature jump.
Similar Mach numbers from measurements
bolster the likelihood that the
surface brightness edge is due to a shock front.



\section{Summary} \label{sec:diss}
We have presented results from deep 227 ks $\chandra$ 
observations of an early-stage merger A98N/A98S. 
Our analysis indicates that the cluster is experiencing a 
complex merger. 
While the northern sub-cluster (A98N) 
and southern sub-cluster (A98S) are merging 
along the N-S, the two sub-clusters of 
A98S (A98Sa and A98Sb) are undergoing a 
later-stage merger in the E-W direction. 
We summarize our findings below -
\begin{itemize}
    \item The residual image of A98N shown in Figure 
\ref{fig:ggm_img}
reveals a gas sloshing spiral winding clockwise 
when traced inward from a large radius.
The sloshing spiral in A98N
indicates that it too experienced 
a separate merger, probably in the
past few Gyr based on the timescale 
for the formation of the sloshing
spirals seen in simulations.

\item We have detected two cold fronts in the north and 
east directions at about 60 kpc and 70 kpc, 
respectively, from the A98N core. 
For the northern cold front, we obtain a drop 
in the deprojected electron density by a factor of 
$\rho_2/\rho_1$ $\approx$ 1.47 $\pm$ 0.01. 
This jump corresponds to an increase in the 
projected temperature by a factor of 1.4. 
Similarly, for the eastern cold fornt, 
we find a decrease in the electron density 
by 
a factor of $\rho_2/\rho_1$ $\approx$ 
1.38 $\pm$ 0.03 and an increment in the 
temperature by a factor of 1.3. 
These two cold fronts are likely to be 
associated with the gas sloshing at the 
central region of A98N.

\item With our deeper $\chandra$ data,
we find the average temperature of the 
A98Sa, $kT$ = 3.70$^{+0.33}_{-0.28}$ keV 
and of the A98Sb, $kT$ = 2.47$^{+0.21}_{-0.20}$ keV, 
consistent with 
that of \citet{2014ApJ...791..104P}. 
We report that the eastern sub-cluster is cooler 
than the western sub-cluster 
with about a 5-$\sigma$ significance.

\item Figure \ref{fig:resid_a98s} 
shows the excess X-ray emission near 
the central region of A98Sb appears to be a 
spiral pattern winding counter-clockwise 
if traced inward from a larger radius. 
The spiral structure is more likely to be 
associated with the gas sloshing at the 
central region of A98Sb, which a recent 
merger may trigger. We have detected a cold front in the 
east direction of A98Sb associated with the 
gas sloshing spiral. 
We measure a drop in the 
deprojected electron density by a 
factor of 
$\rho_2/\rho_1$ $\approx$ 2.8 $\pm$ 0.02
and a rise in the 
projected
gas temperature 
by a factor of 3.6 across the front.

\item The unsharp-masked image shown in 
Figure \ref{fig:A98S_unsharp_resid}
exhibits 
a tail of X-ray emission in the 
south/southwest
direction
of the A98Sb core. 
Our measurement shows that the tail is 
considerably cooler (1.8$^{+0.3}_{-0.2}$ keV) 
than the surrounding gas 
(3.1$^{+0.6}_{-0.4}$ keV) 
at a 4.2-$\sigma$ level and the A98Sb core 
(2.47$^{+0.21}_{-0.20}$ keV) at a 2.9-$\sigma$ level. 
This implies that the tail is likely to be 
ram pressure stripped material from a cool-core remnant.

\item We have reported detecting a surface brightness 
edge at 375 kpc southwest from the A98Sa core 
visible in Figure \ref{fig:A98sa_west_temp}. 
The deprojected 
electron density drops by a factor 
of $\rho_{2}/\rho_{1}$ $\approx$ 
1.7 $\pm$ 0.5 and the projected
temperature drops by a factor of 
$\approx$ 1.5, suggesting the edge 
is a shock front with a 
Mach number of $\cal{M}$ $\approx$ 1.5.


\end{itemize}

\begin{acknowledgments}
This work is based on observations
obtained with $\chandra$ observatory, a NASA mission.
AS and SR are
supported by the grant from NASA's  
Chandra X-ray Observatory, 
grant number GO9-20112X.
CLS was supported in part by 
NASA $\xmm$ grant 
80NSSC22K1510 and 
$\chandra$ grant GO1-22120X.
\end{acknowledgments}

\bibliography{sample631}{}

\begin{thebibliography}{}
\expandafter\ifx\csname natexlab\endcsname\relax\def\natexlab#1{#1}\fi
\providecommand{\url}[1]{\href{#1}{#1}}
\providecommand{\dodoi}[1]{doi:~\href{http://doi.org/#1}{\nolinkurl{#1}}}
\providecommand{\doeprint}[1]{\href{http://ascl.net/#1}{\nolinkurl{http://ascl.net/#1}}}
\providecommand{\doarXiv}[1]{\href{https://arxiv.org/abs/#1}{\nolinkurl{https://arxiv.org/abs/#1}}}

\bibitem[{{Abell} {et~al.}(1989){Abell}, {Corwin}, \&
  {Olowin}}]{1989ApJS...70....1A}
{Abell}, G.~O., {Corwin}, Harold~G., J., \& {Olowin}, R.~P. 1989, \apjs, 70, 1,
  \dodoi{10.1086/191333}

\bibitem[{{Alvarez} {et~al.}(2022){Alvarez}, {Randall}, {Su}, {Sarkar},
  {Walker}, {Lee}, {Sarazin}, \& {Blanton}}]{2022ApJ...938...51A}
{Alvarez}, G.~E., {Randall}, S.~W., {Su}, Y., {et~al.} 2022, \apj, 938, 51,
  \dodoi{10.3847/1538-4357/ac91d3}

\bibitem[{{Ascasibar} \& {Markevitch}(2006)}]{2006ApJ...650..102A}
{Ascasibar}, Y., \& {Markevitch}, M. 2006, \apj, 650, 102,
  \dodoi{10.1086/506508}

\bibitem[{{Asplund} {et~al.}(2009){Asplund}, {Grevesse}, {Sauval}, \&
  {Scott}}]{2009ARA&A..47..481A}
{Asplund}, M., {Grevesse}, N., {Sauval}, A.~J., \& {Scott}, P. 2009, \araa, 47,
  481, \dodoi{10.1146/annurev.astro.46.060407.145222}

\bibitem[{{Blanton} {et~al.}(2011){Blanton}, {Randall}, {Clarke}, {Sarazin},
  {McNamara}, {Douglass}, \& {McDonald}}]{2011ApJ...737...99B}
{Blanton}, E.~L., {Randall}, S.~W., {Clarke}, T.~E., {et~al.} 2011, \apj, 737,
  99, \dodoi{10.1088/0004-637X/737/2/99}

\bibitem[{{Botteon} {et~al.}(2018){Botteon}, {Gastaldello}, \&
  {Brunetti}}]{2018MNRAS.476.5591B}
{Botteon}, A., {Gastaldello}, F., \& {Brunetti}, G. 2018, \mnras, 476, 5591,
  \dodoi{10.1093/mnras/sty598}

\bibitem[{{Burns} {et~al.}(1994){Burns}, {Rhee}, {Owen}, \&
  {Pinkney}}]{1994ApJ...423...94B}
{Burns}, J.~O., {Rhee}, G., {Owen}, F.~N., \& {Pinkney}, J. 1994, \apj, 423,
  94, \dodoi{10.1086/173792}

\bibitem[{{Cash}(1979)}]{1979ApJ...228..939C}
{Cash}, W. 1979, \apj, 228, 939, \dodoi{10.1086/156922}

\bibitem[{{Clarke} {et~al.}(2019){Clarke}, {Scaife}, {Shimwell}, {van Weeren},
  {Bonafede}, {Heald}, {Brunetti}, {Cantwell}, {de Gasperin}, {Br{\"u}ggen},
  {Botteon}, {Hoeft}, {Horellou}, {Cassano}, {Harwood}, \&
  {R{\"o}ttgering}}]{2019A&A...627A.176C}
{Clarke}, A.~O., {Scaife}, A.~M.~M., {Shimwell}, T., {et~al.} 2019, \aap, 627,
  A176, \dodoi{10.1051/0004-6361/201935584}

\bibitem[{{Clarke} {et~al.}(2004){Clarke}, {Blanton}, \&
  {Sarazin}}]{2004ApJ...616..178C}
{Clarke}, T.~E., {Blanton}, E.~L., \& {Sarazin}, C.~L. 2004, \apj, 616, 178,
  \dodoi{10.1086/424911}

\bibitem[{{Douglass} {et~al.}(2011){Douglass}, {Blanton}, {Clarke}, {Randall},
  \& {Wing}}]{2011ApJ...743..199D}
{Douglass}, E.~M., {Blanton}, E.~L., {Clarke}, T.~E., {Randall}, S.~W., \&
  {Wing}, J.~D. 2011, \apj, 743, 199, \dodoi{10.1088/0004-637X/743/2/199}

\bibitem[{{Douglass} {et~al.}(2008){Douglass}, {Blanton}, {Clarke}, {Sarazin},
  \& {Wise}}]{2008ApJ...673..763D}
{Douglass}, E.~M., {Blanton}, E.~L., {Clarke}, T.~E., {Sarazin}, C.~L., \&
  {Wise}, M. 2008, \apj, 673, 763, \dodoi{10.1086/523886}

\bibitem[{{Douglass} {et~al.}(2018){Douglass}, {Blanton}, {Randall}, {Clarke},
  {Edwards}, {Sabry}, \& {ZuHone}}]{2018ApJ...868..121D}
{Douglass}, E.~M., {Blanton}, E.~L., {Randall}, S.~W., {et~al.} 2018, \apj,
  868, 121, \dodoi{10.3847/1538-4357/aae9e7}

\bibitem[{{Erdim} \& {Hudaverdi}(2019)}]{2019MNRAS.488.2917E}
{Erdim}, M.~K., \& {Hudaverdi}, M. 2019, \mnras, 488, 2917,
  \dodoi{10.1093/mnras/stz1912}

\bibitem[{{Fabian} {et~al.}(2003){Fabian}, {Sanders}, {Allen}, {Crawford},
  {Iwasawa}, {Johnstone}, {Schmidt}, \& {Taylor}}]{2003MNRAS.344L..43F}
{Fabian}, A.~C., {Sanders}, J.~S., {Allen}, S.~W., {et~al.} 2003, \mnras, 344,
  L43, \dodoi{10.1046/j.1365-8711.2003.06902.x}

\bibitem[{{Fabian} {et~al.}(2001){Fabian}, {Sanders}, {Ettori}, {Taylor},
  {Allen}, {Crawford}, {Iwasawa}, \& {Johnstone}}]{2001MNRAS.321L..33F}
{Fabian}, A.~C., {Sanders}, J.~S., {Ettori}, S., {et~al.} 2001, \mnras, 321,
  L33, \dodoi{10.1046/j.1365-8711.2001.04243.x}

\bibitem[{{Fabian} {et~al.}(2011){Fabian}, {Sanders}, {Allen}, {Canning},
  {Churazov}, {Crawford}, {Forman}, {Gabany}, {Hlavacek-Larrondo}, {Johnstone},
  {Russell}, {Reynolds}, {Salom{\'e}}, {Taylor}, \&
  {Young}}]{2011MNRAS.418.2154F}
{Fabian}, A.~C., {Sanders}, J.~S., {Allen}, S.~W., {et~al.} 2011, \mnras, 418,
  2154, \dodoi{10.1111/j.1365-2966.2011.19402.x}

\bibitem[{{Forman} {et~al.}(1981){Forman}, {Bechtold}, {Blair}, {Giacconi},
  {van Speybroeck}, \& {Jones}}]{1981ApJ...243L.133F}
{Forman}, W., {Bechtold}, J., {Blair}, W., {et~al.} 1981, \apjl, 243, L133,
  \dodoi{10.1086/183459}

\bibitem[{{Gastaldello} {et~al.}(2013){Gastaldello}, {Di Gesu}, {Ghizzardi},
  {Giacintucci}, {Girardi}, {Roediger}, {Rossetti}, {Brighenti}, {Buote},
  {Eckert}, {Ettori}, {Humphrey}, \& {Mathews}}]{2013ApJ...770...56G}
{Gastaldello}, F., {Di Gesu}, L., {Ghizzardi}, S., {et~al.} 2013, \apj, 770,
  56, \dodoi{10.1088/0004-637X/770/1/56}

\bibitem[{{Ge} {et~al.}(2020){Ge}, {Liu}, {Sun}, {Yu}, {Rudnick}, {Eilek},
  {Owen}, {Dasadia}, {Rossetti}, {Markevitch}, {Clarke}, {Jones}, {Ghizzardi},
  {Venturi}, {Finoguenov}, \& {Eckert}}]{2020MNRAS.497.4704G}
{Ge}, C., {Liu}, R.-Y., {Sun}, M., {et~al.} 2020, \mnras, 497, 4704,
  \dodoi{10.1093/mnras/staa2320}

\bibitem[{{G{\'o}mez} \& {Calder{\'o}n}(2020)}]{2020AJ....160..152G}
{G{\'o}mez}, P.~L., \& {Calder{\'o}n}, D. 2020, \aj, 160, 152,
  \dodoi{10.3847/1538-3881/aba831}

\bibitem[{{Hu} {et~al.}(2021){Hu}, {Xu}, {Zhu}, {Shan}, {Zhu}, {Fan}, {Zhao},
  {Liu}, {Siew}, {Zhang}, {Gu}, {Johnston-Hollitt}, {Kang}, {Tan}, {Chang}, \&
  {Wu}}]{2021ApJ...913....8H}
{Hu}, D., {Xu}, H., {Zhu}, Z., {et~al.} 2021, \apj, 913, 8,
  \dodoi{10.3847/1538-4357/abf09e}

\bibitem[{{Ichinohe} {et~al.}(2019){Ichinohe}, {Simionescu}, {Werner},
  {Fabian}, \& {Takahashi}}]{2019MNRAS.483.1744I}
{Ichinohe}, Y., {Simionescu}, A., {Werner}, N., {Fabian}, A.~C., \&
  {Takahashi}, T. 2019, \mnras, 483, 1744, \dodoi{10.1093/mnras/sty3257}

\bibitem[{{Ichinohe} {et~al.}(2021){Ichinohe}, {Simionescu}, {Werner},
  {Markevitch}, \& {Wang}}]{2021MNRAS.504.2800I}
{Ichinohe}, Y., {Simionescu}, A., {Werner}, N., {Markevitch}, M., \& {Wang},
  Q.~H.~S. 2021, \mnras, 504, 2800, \dodoi{10.1093/mnras/stab1060}

\bibitem[{{Ignesti} {et~al.}(2018){Ignesti}, {Gitti}, {Brunetti}, {O'Sullivan},
  {Sarazin}, \& {Wong}}]{2018A&A...610A..89I}
{Ignesti}, A., {Gitti}, M., {Brunetti}, G., {et~al.} 2018, \aap, 610, A89,
  \dodoi{10.1051/0004-6361/201731380}

\bibitem[{{Ignesti} {et~al.}(2020){Ignesti}, {Shimwell}, {Brunetti}, {Gitti},
  {Intema}, {van Weeren}, {Hardcastle}, {Clarke}, {Botteon}, {Di Gennaro},
  {Br{\"u}ggen}, {Browne}, {Mandal}, {R{\"o}ttgering}, {Cuciti}, {de Gasperin},
  {Cassano}, \& {Scaife}}]{2020A&A...643A.172I}
{Ignesti}, A., {Shimwell}, T., {Brunetti}, G., {et~al.} 2020, \aap, 643, A172,
  \dodoi{10.1051/0004-6361/202039056}

\bibitem[{{Johnson} {et~al.}(2010){Johnson}, {Markevitch}, {Wegner}, {Jones},
  \& {Forman}}]{2010ApJ...710.1776J}
{Johnson}, R.~E., {Markevitch}, M., {Wegner}, G.~A., {Jones}, C., \& {Forman},
  W.~R. 2010, \apj, 710, 1776, \dodoi{10.1088/0004-637X/710/2/1776}

\bibitem[{{Jones} \& {Forman}(1999)}]{1999ApJ...511...65J}
{Jones}, C., \& {Forman}, W. 1999, \apj, 511, 65, \dodoi{10.1086/306646}

\bibitem[{{Kalberla} {et~al.}(2005){Kalberla}, {Burton}, {Hartmann}, {Arnal},
  {Bajaja}, {Morras}, \& {P{\"o}ppel}}]{2005A&A...440..775K}
{Kalberla}, P.~M.~W., {Burton}, W.~B., {Hartmann}, D., {et~al.} 2005, \aap,
  440, 775, \dodoi{10.1051/0004-6361:20041864}

\bibitem[{{Kokotanekov} {et~al.}(2018){Kokotanekov}, {Wise}, {de Vries}, \&
  {Intema}}]{2018A&A...618A.152K}
{Kokotanekov}, G., {Wise}, M.~W., {de Vries}, M., \& {Intema}, H.~T. 2018,
  \aap, 618, A152, \dodoi{10.1051/0004-6361/201833222}

\bibitem[{{Markevitch} {et~al.}(2002){Markevitch}, {Gonzalez}, {David},
  {Vikhlinin}, {Murray}, {Forman}, {Jones}, \& {Tucker}}]{2002ApJ...567L..27M}
{Markevitch}, M., {Gonzalez}, A.~H., {David}, L., {et~al.} 2002, \apjl, 567,
  L27, \dodoi{10.1086/339619}

\bibitem[{{Markevitch} \& {Vikhlinin}(2007)}]{2007PhR...443....1M}
{Markevitch}, M., \& {Vikhlinin}, A. 2007, \physrep, 443, 1,
  \dodoi{10.1016/j.physrep.2007.01.001}

\bibitem[{{Markevitch} {et~al.}(2001){Markevitch}, {Vikhlinin}, \&
  {Mazzotta}}]{2001ApJ...562L.153M}
{Markevitch}, M., {Vikhlinin}, A., \& {Mazzotta}, P. 2001, \apjl, 562, L153,
  \dodoi{10.1086/337973}

\bibitem[{{Markevitch} {et~al.}(2000){Markevitch}, {Ponman}, {Nulsen}, {Bautz},
  {Burke}, {David}, {Davis}, {Donnelly}, {Forman}, {Jones}, {Kaastra},
  {Kellogg}, {Kim}, {Kolodziejczak}, {Mazzotta}, {Pagliaro}, {Patel}, {Van
  Speybroeck}, {Vikhlinin}, {Vrtilek}, {Wise}, \& {Zhao}}]{2000ApJ...541..542M}
{Markevitch}, M., {Ponman}, T.~J., {Nulsen}, P.~E.~J., {et~al.} 2000, \apj,
  541, 542, \dodoi{10.1086/309470}

\bibitem[{{Mathews} \& {Brighenti}(2008)}]{2008ApJ...685..128M}
{Mathews}, W.~G., \& {Brighenti}, F. 2008, \apj, 685, 128,
  \dodoi{10.1086/590402}

\bibitem[{{Nulsen} {et~al.}(2013){Nulsen}, {Li}, {Forman}, {Kraft}, {Lal},
  {Jones}, {Zhuravleva}, {Churazov}, {Sanders}, {Fabian}, {Johnson}, \&
  {Murray}}]{2013ApJ...775..117N}
{Nulsen}, P. E.~J., {Li}, Z., {Forman}, W.~R., {et~al.} 2013, \apj, 775, 117,
  \dodoi{10.1088/0004-637X/775/2/117}

\bibitem[{{O'Donoghue} {et~al.}(1993){O'Donoghue}, {Eilek}, \&
  {Owen}}]{1993ApJ...408..428O}
{O'Donoghue}, A.~A., {Eilek}, J.~A., \& {Owen}, F.~N. 1993, \apj, 408, 428,
  \dodoi{10.1086/172600}

\bibitem[{{Paterno-Mahler} {et~al.}(2013){Paterno-Mahler}, {Blanton},
  {Randall}, \& {Clarke}}]{2013ApJ...773..114P}
{Paterno-Mahler}, R., {Blanton}, E.~L., {Randall}, S.~W., \& {Clarke}, T.~E.
  2013, \apj, 773, 114, \dodoi{10.1088/0004-637X/773/2/114}

\bibitem[{{Paterno-Mahler} {et~al.}(2014){Paterno-Mahler}, {Randall}, {Bulbul},
  {Andrade-Santos}, {Blanton}, {Jones}, {Murray}, \&
  {Johnson}}]{2014ApJ...791..104P}
{Paterno-Mahler}, R., {Randall}, S.~W., {Bulbul}, E., {et~al.} 2014, \apj, 791,
  104, \dodoi{10.1088/0004-637X/791/2/104}

\bibitem[{{Randall} {et~al.}(2008){Randall}, {Nulsen}, {Forman}, {Jones},
  {Machacek}, {Murray}, \& {Maughan}}]{2008ApJ...688..208R}
{Randall}, S., {Nulsen}, P., {Forman}, W.~R., {et~al.} 2008, \apj, 688, 208,
  \dodoi{10.1086/592324}

\bibitem[{{Randall} {et~al.}(2009){Randall}, {Jones}, {Kraft}, {Forman}, \&
  {O'Sullivan}}]{2009ApJ...696.1431R}
{Randall}, S.~W., {Jones}, C., {Kraft}, R., {Forman}, W.~R., \& {O'Sullivan},
  E. 2009, \apj, 696, 1431, \dodoi{10.1088/0004-637X/696/2/1431}

\bibitem[{{Roediger} {et~al.}(2011){Roediger}, {Br{\"u}ggen}, {Simionescu},
  {B{\"o}hringer}, {Churazov}, \& {Forman}}]{2011MNRAS.413.2057R}
{Roediger}, E., {Br{\"u}ggen}, M., {Simionescu}, A., {et~al.} 2011, \mnras,
  413, 2057, \dodoi{10.1111/j.1365-2966.2011.18279.x}

\bibitem[{{Roediger} {et~al.}(2012){Roediger}, {Lovisari}, {Dupke},
  {Ghizzardi}, {Br{\"u}ggen}, {Kraft}, \& {Machacek}}]{2012MNRAS.420.3632R}
{Roediger}, E., {Lovisari}, L., {Dupke}, R., {et~al.} 2012, \mnras, 420, 3632,
  \dodoi{10.1111/j.1365-2966.2011.20287.x}

\bibitem[{{Rossetti} {et~al.}(2013){Rossetti}, {Eckert}, {De Grandi},
  {Gastaldello}, {Ghizzardi}, {Roediger}, \& {Molendi}}]{2013A&A...556A..44R}
{Rossetti}, M., {Eckert}, D., {De Grandi}, S., {et~al.} 2013, \aap, 556, A44,
  \dodoi{10.1051/0004-6361/201321319}

\bibitem[{{Russell} {et~al.}(2010){Russell}, {Sanders}, {Fabian}, {Baum},
  {Donahue}, {Edge}, {McNamara}, \& {O'Dea}}]{2010MNRAS.406.1721R}
{Russell}, H.~R., {Sanders}, J.~S., {Fabian}, A.~C., {et~al.} 2010, \mnras,
  406, 1721, \dodoi{10.1111/j.1365-2966.2010.16822.x}

\bibitem[{{Sarkar} {et~al.}(2022){Sarkar}, {Randall}, {Su}, {Alvarez},
  {Sarazin}, {Nulsen}, {Blanton}, {Forman}, {Jones}, {Bulbul}, {Zuhone},
  {Andrade-Santos}, {Johnson}, \& {Chakraborty}}]{2022ApJ...935L..23S}
{Sarkar}, A., {Randall}, S., {Su}, Y., {et~al.} 2022, \apjl, 935, L23,
  \dodoi{10.3847/2041-8213/ac86d4}

\bibitem[{{Smith} {et~al.}(2001){Smith}, {Brickhouse}, {Liedahl}, \&
  {Raymond}}]{2001ApJ...556L..91S}
{Smith}, R.~K., {Brickhouse}, N.~S., {Liedahl}, D.~A., \& {Raymond}, J.~C.
  2001, \apjl, 556, L91, \dodoi{10.1086/322992}

\bibitem[{{Su} {et~al.}(2017){Su}, {Nulsen}, {Kraft}, {Roediger}, {ZuHone},
  {Jones}, {Forman}, {Sheardown}, {Irwin}, \& {Randall}}]{2017ApJ...851...69S}
{Su}, Y., {Nulsen}, P. E.~J., {Kraft}, R.~P., {et~al.} 2017, \apj, 851, 69,
  \dodoi{10.3847/1538-4357/aa989e}

\bibitem[{{Vikhlinin} {et~al.}(2001){Vikhlinin}, {Markevitch}, \&
  {Murray}}]{2001ApJ...551..160V}
{Vikhlinin}, A., {Markevitch}, M., \& {Murray}, S.~S. 2001, \apj, 551, 160,
  \dodoi{10.1086/320078}

\bibitem[{{Walker} {et~al.}(2017){Walker}, {Hlavacek-Larrondo},
  {Gendron-Marsolais}, {Fabian}, {Intema}, {Sanders}, {Bamford}, \& {van
  Weeren}}]{2017MNRAS.468.2506W}
{Walker}, S.~A., {Hlavacek-Larrondo}, J., {Gendron-Marsolais}, M., {et~al.}
  2017, \mnras, 468, 2506, \dodoi{10.1093/mnras/stx640}

\bibitem[{{Walker} {et~al.}(2020){Walker}, {Mirakhor}, {ZuHone}, {Sanders},
  {Fabian}, \& {Diwanji}}]{2020arXiv200614043W}
{Walker}, S.~A., {Mirakhor}, M.~S., {ZuHone}, J., {et~al.} 2020, arXiv
  e-prints, arXiv:2006.14043.
\newblock \doarXiv{2006.14043}

\bibitem[{{ZuHone} {et~al.}(2010){ZuHone}, {Markevitch}, \&
  {Johnson}}]{2010ApJ...717..908Z}
{ZuHone}, J.~A., {Markevitch}, M., \& {Johnson}, R.~E. 2010, \apj, 717, 908,
  \dodoi{10.1088/0004-637X/717/2/908}

\bibitem[{{ZuHone} {et~al.}(2011){ZuHone}, {Markevitch}, \&
  {Lee}}]{2011ApJ...743...16Z}
{ZuHone}, J.~A., {Markevitch}, M., \& {Lee}, D. 2011, \apj, 743, 16,
  \dodoi{10.1088/0004-637X/743/1/16}

\end{thebibliography}
\bibliographystyle{aasjournal}



\end{document}